\documentclass[12pt,english]{article}
\usepackage[LGR,T1]{fontenc}
\usepackage[latin2]{inputenc}
\usepackage[letterpaper]{geometry}
\usepackage{color}
\usepackage{amsmath}
\usepackage{amssymb}
\usepackage{graphicx}
\usepackage{setspace}
\makeatletter

\DeclareRobustCommand{\greektext}{%
  \fontencoding{LGR}\selectfont\def\encodingdefault{LGR}}
\DeclareRobustCommand{\textgreek}[1]{\leavevmode{\greektext #1}}
\ProvideTextCommand{\~}{LGR}[1]{\char126#1}

\usepackage{cite}

\makeatother

\usepackage{babel}
\begin{document}
\begin{center}
\textbf{\LARGE{}Light-induced photodissociation on the lowest three
electronic states of $\mathrm{NaH}$ molecule}{\LARGE\par}
\par\end{center}

\begin{center}
{\large{}Otabek Umarov (1,2), András Csehi (1), Péter Badankó (1)
Gábor J. Halász (3) and Ágnes Vibók (1,4)}\\
(1) Department of Theoretical Physics, Doctoral School of Physics,
University of Debrecen, \\
H-4002 Debrecen, PO Box 400, Hungary\\
(2) Department of Optics and Spectroscopy, Samarkand State University,
\\
140104, University blv. 15, Samarkand, Uzbekistan\\
(3) Department of Information Technology, University of Debrecen,
\\
H-4002 Debrecen, PO Box 400 Hungary\\
(4) ELI-ALPS, ELI-HU Non-Profit Ltd, \\
H-6720 Szeged, Dugonics tér 13, Hungary\\
\par\end{center}
\begin{abstract}
It has been known that electronic conical intersections in a molecular
system can also be created by laser light even in diatomics. The direct
consequence of these light-induced degeneracies is the appearance
of a strong mixing between the electronic and vibrational motions,
which has a strong fingerprint on the ultrafast nuclear dynamics.
In the present work, pump and probe numerical simulations have been
performed with the $\mathrm{NaH}$ molecule involving the first three
singlet electronic states ($\mathrm{X^{1}\mathrm{\Sigma}^{+}(X)},$
$\mathrm{A^{1}\mathrm{\Sigma}^{+}(A)}$ and $\mathrm{B^{1}\Pi(B)}$)
and several light-induced degeneracies in the numerical description.
To demonstrate the impact of the multiple light-induced non-adiabatic
effects together with the molecular rotation on the dynamical properties
of the molecule, the dissociation probabilities, kinetic energy release
spectra (KER) and the angular distributions of the photofragments
were calculated by discussing the role of the permanent dipole moment
as well. 

\end{abstract}

\part*{I. Introduction}

Understanding the behavior of atoms and molecules under the effect
of strong electromagnetic fields is an extensively investigated research
field. There are a large number of theoretical and experimental works
available which have been devoted to discuss numerous new phenomena
of light-matter interaction. Although many of these works treat the
dynamical problem of diatomics, starting investigations from the simplest
hydrogen-like ions or molecules to systems with large number of electrons
\cite{Bandrauk1,Bandrauk2,Takasuka1,Takasuka2,Tiwari0,Foudil0,Takasuka3,Tiwari1,Attila1,Zhaopeng1,Foudil1,Attila2,Attila3,Zhaopeng2,Zhang1,Zhang2}.
Nevertheless, several other relevant papers tackle the problem of
photodissociation and fragmentation of polyatomics as well \cite{Banares2,Banares3,Banares4,Banares5,Fabien1,Fabien2,Fabien3,Graham2,Graham3,Graham4,Ignacio1,Ignacio2,Ignacio3,Ignacio4,Ignacio5,Weinacht1,Weinacht2,Weinacht3,Bende1,Bende2}. 

Molecular dissociation is usually treated within the Born-Oppenheimer
(BO) framework, which relies on the separation of the motions of the
electrons and nuclei due to different time scale of their motion.
In most cases, this approach works well and provides an acceptable
treatment of dynamical processes. But at certain nuclear configurations,
in particular, in the close vicinity of degeneracy points or conical
intersections (CIs), the energy exchange between electrons and nuclei
becomes significant and the BO approximation fails \cite{Born1}.
CIs play an important role in non-adiabatic processes acting as efficient
funnels for ultrafast interstate crossings typically on the femtosecond
time scale \cite{Koppel1,Yarkony1,Graham1,Koppel2,Baer1}. 

It is known that a molecular system must possess at least two independent
nuclear degrees of freedom so as to form a CI. Therefore, having only
one nuclear vibrational mode like in the case of diatomics, CIs can
never appear. Nevertheless, if an additional degree of freedom like
the rotation is associated with the system due to some interactions
with the environment (e.g. laser-molecule interaction), CIs can arise.
It has been found that, by applying external laser field, light-induced
avoided crossings (LIACs) and/or light-induced conical intersections
(LIACs) can be formed even in diatomics \cite{Nimrod1}. In contrast
to the ACs and CIs occurring in nature, the energy at which the LIACs
and LICIs are found, and their non-adiabatic strengths can be controlled
by the laser intensity and frequency, respectively. By changing the
position and structure of these objects, it is possible to manipulate
the non-adiabatic dynamics in molecular systems. 

Detailed theoretical and experimental investigations demonstrate that
the LICIs can have significant impact on several spectroscopic, dynamical
and topological properties of both diatomic \cite{Gabor1,Gabor2,Gabor3,Gabor4,Andris1,Andris2,Andris3,Andris4,Tamas1,Bandrauk3,Buksbaum1,Nature1}
and polyatomic \cite{Buksbaum2,Banares,Csaba1} systems. 

As is known the $\mathrm{NaH}$ molecule has an astrophysical importance
and the photodissociation is one of the channels for the destruction
of this alkali hydride molecule in interstellar clouds. In the present
paper, we perform pump and probe numerical simulations so as to investigate
the non-adiabatic light-induced photodissociation process of the $\mathrm{NaH}$
molecule \cite{NaH1,NaH2,NaH3,NaH4} in the presence of molecular
rotation. The inclusion of the rotation in the dynamical description
makes it possible to take appropriately into account the light-induced
non-adiabatic effect generated by the strong laser field. Two and
one-dimensional numerical simulations are performed, in which, in
addition to the molecular vibration the rotation will also be included
considering it either as a dynamic variable (2D scheme, LICI framework)
or only as a parameter (1D scheme, LIAC framework). Three electronic
states ($\mathrm{X^{1}\mathrm{\Sigma}^{+}(X)},$ $\mathrm{A^{1}\mathrm{\Sigma}^{+}(A)}$
and $\mathrm{B^{1}\Pi(X)}$) are involved in the numerical descriptions
and particular attention will be paid to the proper discussion of
the role of the permanent dipole moment that was completely neglected
before \cite{NaH1,NaH2,NaH3,NaH4}. In order to exhaustively describe
the light-induced non-adiabatic dissociation dynamics of this system,
several different dynamical properties like dissociation probabilities,
kinetic energy release spectra (KER), as well as fragment angular
distributions have been studied. 

The article is structured as follows. In the next section, the methodology
and algorithms required for the theoretical study are presented. Namely,
the working Hamiltonian, the applied electric fields, the calculated
dynamical quantities as well as the numerical details are briefly
summarized. In the third section, we present and discuss the numerical
results for the $\mathrm{NaH}$ system. Finally, in the last part,
conclusions will be given.

\part*{II. Methodology and physical background }

In what follows, a brief summary will be provided concerning the electronic
structure quantities of $\mathrm{\mathrm{NaH}}$. Then the time-dependent
nuclear Hamiltonian, which governs the dynamics of the system will
be defined. In the last two parts of this section a brief description
of the time-dependent electric field of the applied laser pulses,
the theoretical approach used to compute the time evolution of the
nuclear wave packet, as well as the expressions of the computed dynamical
quantities are given. 

\part*{A. The sodium hydride molecule}

In our numerical simulations, the $\mathrm{NaH}$ molecule is described
as a three-level system involving the three lowest-lying singlet electronic
states, namely $\mathrm{X^{1}\mathrm{\Sigma}^{+}(X)}$ , $\mathrm{A^{1}\mathrm{\Sigma}^{+}(A)}$
and $\mathrm{B^{1}\Pi(B)}$ (labeled throughout the paper as $\mathrm{V_{X}}(\mathrm{R)}$,
$\mathrm{V_{A}}(\mathrm{R)}$ and $\mathrm{V_{B}}(\mathrm{R)}$).
The corresponding potential energy curves are visualized in Fig. 1a.
Figures 1b and 1c present the transition ($\overrightarrow{\mu}_{i,j}(R)=\left\langle \phi_{i}\mid\mathrm{\Sigma}_{k}\overrightarrow{r_{k}}\mid\phi_{j}\right\rangle $)
and the permanent ($\overrightarrow{\mu}_{i}(R)=\left\langle \phi_{i}\mid\mathrm{\Sigma}_{k}\overrightarrow{r_{k}}\mid\phi_{i}\right\rangle $)
dipole moments (TDM and PDM) ($i,j\in\left(\mathrm{X},\mathrm{A},\mathrm{B}\right)),$
respectively. It should be noted that all the permanent dipole moments
and the transition dipole moment $\overrightarrow{\mu}_{X,A}$ are
parallel with the molecular axis, while the ones corresponding to
the $\Sigma-\Pi$ transition ($\overrightarrow{\mu}_{X,B}$ and $\overrightarrow{\mu}_{A,B}$)
are perpendicular. Numerical computation of these electronic structure
quantities has been carried out by using the MOLPRO program package
\cite{Molpro} at the MRCI/CAS(2/9)/aug-cc-pV5Z level of theory. The
number of active electrons and MOs in the individual irreducible representations
of the $\mathrm{C_{2v}}$ point group were $\mathrm{A}\rightarrow\mathrm{2/5,B_{1}}\rightarrow\mathrm{0/2,B_{2}}\rightarrow\mathrm{0/2,A_{2}}\rightarrow\mathrm{0/0}$.
By using these values we could achieve reasonable agreement with other
results in the literature \cite{JCP1,JCP2}.

\begin{figure}
\begin{centering}
\includegraphics[width=0.48\textwidth]{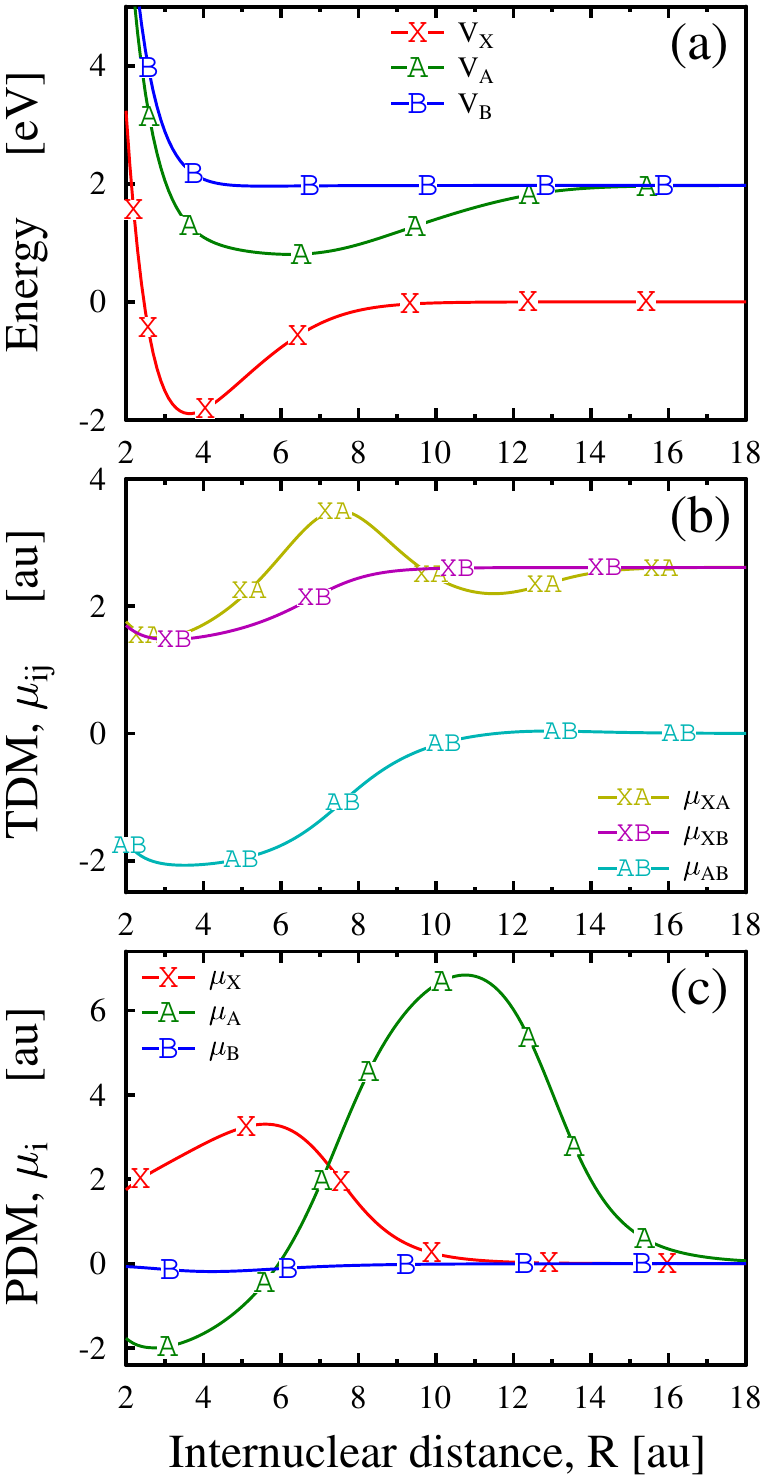}
\par\end{centering}
\caption{\label{fig:fig1} (a) Lowest three adiabatic potential energy curves
of the $\mathrm{NaH}$ molecule. (b) Transition dipole moment (TDM)
functions between the different adiabatic electronic states. (c) Permanent
dipole moment (PDM) functions of the three adiabatic electronic states. }
\end{figure}

\part*{B. The working Hamiltonian}

In the space of the three lowest singlet electronic states the following
full time-dependent nuclear Hamiltonian of $\mathrm{NaH}$ holds for
the rovibronic nuclear motions: 

\begin{align}
\mathrm{H} & =\mathrm{(-\frac{1}{2M_{r}}\frac{\partial^{2}}{\partial R^{2}}+\frac{L_{\theta}^{2}}{2M_{r}R^{2}}).\mathbf{1}+\left(\begin{array}{ccc}
\mathrm{V_{X}(}\mathrm{R}) & 0 & 0\\
0 & \mathrm{V_{A}}(\mathrm{R}) & 0\\
0 & 0 & \mathrm{V_{B}}(\mathrm{R})
\end{array}\right)}+\label{eq:Hamilton}\\
-\mathrm{E(t)} & \left(\mathrm{\begin{array}{ccc}
\mu_{X}\cos(\theta) & \mu_{X,A}\cos(\theta) & \mu_{X,B}\sin(\theta)\\
\mu_{A,X}\cos(\theta) & \mu_{A}\cos(\theta) & \mu_{A,B}\sin(\theta)\\
\mu_{B,X}\sin(\theta) & \mu_{B,A}\sin(\theta) & \mu_{B}\cos(\theta)
\end{array}}\right).\nonumber 
\end{align}
Here the first term in H describes the kinetic energy of the system
with $\mathrm{R}$ and $\mathrm{\mathrm{\theta}}$ being the molecular
vibrational and rotational coordinates, respectively. $\mathbf{1}$
is the three-dimensional unit matrix, $\mathrm{\mathrm{M_{r}}}$ is
the reduced mass, and $\mathrm{L_{\theta}}$ is the angular momentum
operator of the nuclei with $\mathrm{m=0}$. The rotational coordinate
$\mathrm{\theta}$, represents the angle between the internuclear
axis and the laser polarization direction. $\mathrm{V_{X}(R)}$, $\mathrm{V_{A}(R)}$
and $\mathrm{\mathrm{V_{B}(R)}}$ are the first three singlet electronic
states. The third term represents the light-matter interaction in
dipole approximation. $\mathrm{E\left(t\right)}$ characterizes the
time-dependent electric field and the dipole matrix contains the permanent
as well as the transition dipole moments, as these were defined in
the former section. Atomic units ($\mathrm{e=m_{e}=\hbar=1)}$ are
used throughout the article. Expression given by Eq. (1) will be used
throughout the paper, as the working Hamiltonian. 

\section*{C. The applied electric field}

The linearly polarized $\mathrm{E(t)}$ laser field applied in the
simulations is the sum of two components:

\begin{equation}
\mathrm{E(t)}=\mathrm{E_{pm}f_{pm}(t)\cos(\omega_{pm}t)+E_{pr}f_{pr}(t)\cos(\omega_{pr}t)}.\label{eq:laser field}
\end{equation}

The first term represents a pump pulse with energy $\hbar\omega_{pm}$
and amplitude $E_{pm}$ which initiates the dynamics of the system
by transferring some amounts of the initial ground $\mathrm{V_{X}(R)}$
state population to the excited $\mathrm{V_{A}(R)}$ state. The second
term provides a probe pulse with energy $\hbar\omega_{pr}$ and amplitude
$E_{pr}$ transferring further the population from the $\mathrm{V_{A}(R)}$
state to the other two $\mathrm{V_{X}(R)}$ and $\mathrm{V_{B}(R)}$
dissociation channels. Cos-square-shaped pulses were used for the
envelope functions of the pump $\mathrm{f_{pm}(t)}$ and probe $\mathrm{f_{pr}(t)}$
pulses: 

\begin{equation}
\mathrm{f_{pm}(t)=\cos^{2}\left(\frac{\pi(t-t_{pm})}{\tau_{pm}}\right)}\label{eq:envelop1}
\end{equation}

\begin{equation}
\mathrm{f_{pr}(t)=\cos^{2}\left(\frac{\pi(t-t_{pr})}{\tau_{pr}}\right)}.\label{eq:envelop2}
\end{equation}

Here $\mathrm{t_{pm}=0\,\mathrm{fs}}\mathrm{}$ and $\mathrm{t_{pr}=0...1000\,\mathrm{fs}}$
are the centers, while $\mathrm{\tau_{pm}}$ and $\mathrm{\tau_{pr}}$
are the durations of the pump and probe laser pulses, respectively.
One can define the delay time as: $\mathrm{t_{del}=t_{pr}-t_{pm}}$.
During the numerical simulations the energy, the intensity and the
duration of the pump and probe pulses were fixed. For the pump pulse
an efficient population was provided by the parameters $\mathrm{\omega_{pm}=3.24\,\mathrm{eV}}$,
$\mathrm{I_{pm}}=$$1\cdot10^{13}\,\mathrm{W/cm^{2}}$ and $\mathrm{\tau_{pm}=8\,fs}$,
while for the probe pulse the $\mathrm{\mathrm{\omega_{pr}}=1.08\,\mathrm{eV}}$,
$\mathrm{I_{pr}=}$$1\cdot10^{12}\mathrm{W/cm^{2}}$ and $\mathrm{\mathrm{\tau_{pr}}=20\,fs}$
parameters were able to couple properly the $\mathrm{V_{A}(R)}$ state
with the $\mathrm{\mathrm{V_{X}(R)}}$ and $\mathrm{V_{B}(R)}$ channels. 

So as to demonstrate the essence of the LICI phenomenon we can borrow
the Floquet framework of the nuclear Hamiltonian (see \cite{Gabor1,Gabor2,Gabor3,Gabor4}
for a detailed discussion). This presentation provides an illustrative
picture and is frequently used to understand different phenomena in
the field of light-matter physics (see in Fig. 2). 

In this picture, the pump laser pulse shifts the energy of the $\mathrm{V_{X}(R)}$
ground potential curve by $\mathrm{\mathrm{\hbar\omega_{pm}}}$ and
a crossing (LICI$\mathrm{_{pump}})$ between the shifted ground ($\mathrm{V_{X}(R)}+\hbar\omega_{pm}$)
and the first excited potential energy $\mathrm{V_{A}(R)}$ curves
is created (see Fig. 2a). This process then initiates a wave packet
dynamics on the state $\mathrm{\mathrm{V_{A}(R)}}$ which can be tested
by a probe laser pulse which couples the $\mathrm{V_{A}(R)}$ both
to the $\mathrm{V_{B}(R)}$ as well as to the $\mathrm{V_{X}(R)}$
potential energy states. The action of the probe pulse can be described
in the Floquet framework, as the $\mathrm{V_{A}(R)}$ first excited
potential curve being shifted upwards and downwards by $\mathrm{\mathrm{\hbar\omega_{pm}}}$.
Crossings between the upwards shifted first excited ($\mathrm{V_{A}(R)}+\hbar\omega_{pr}$)
and the second excited $\mathrm{V_{B}(R)}$ potential energy curves,
as well as the downwards shifted first excited ($\mathrm{V_{A}(R)}-\hbar\omega_{pr}$)
and the ground $\mathrm{V_{X}(R)}$ potential energy curves are created,
respectively. We note that the pump pulse has already created a LICI$\mathrm{_{pump}}$
between the shifted ground ($\mathrm{V_{X}(R)}+\hbar\omega_{pm}$)
and the first excited $\mathrm{V_{A}(R)}$ potential energy curves
(see Fig. 2a), and its effect is manifested in the initiation of the
wave packet rotation dynamics on the $\mathrm{V_{A}(R)}$. By diagonalizing
the corresponding Floquet form of the working Hamiltonian Eq.(\ref{eq:Hamilton}),
one can obtain the light-induced adiabatic potential energy surfaces
which can be seen in Fig. 2b. The corresponding light-induced adiabatic
surfaces cross each other giving rise to light-induced conical intersections
whenever the conditions sin \textgreek{j} = 0(\textgreek{j} = 0, \textgreek{j}
= \textgreek{p}) and $\mathrm{(V_{A}(R)}-\hbar\omega_{pr})=\mathrm{V_{X}(r)}$
as well as cos \textgreek{j} = 0(\textgreek{j} = \textgreek{p}/2)
and $\mathrm{(V_{A}(R)}+\hbar\omega_{pr})=\mathrm{V_{B}(r)}$ are
simultaneously fulfilled. For the parallel transition, the sin \textgreek{j}
factors are replaced by cos \textgreek{j} in Eq.(\ref{eq:Hamilton}).
Namely, for perpendicular transitions, the LICIs occur at \textgreek{j}
= 0; \textgreek{j} = \textgreek{p}, while for parallel transitions,
at \textgreek{j} = \textgreek{p}/2. \textcolor{black}{For those laser
parameters which are applied during the present calculations two LICIs
(LICI1 and LICI2) and one LIAC are formed by the effect of the probe
pulse (see Fig. 2a). }

It is important to emphasize that in contrast to the case provided
by natural CIs of field-free molecules, the LICI is controllable in
the sense that the laser frequency determines its position and energy
in the nuclear configuration space, while the strength of the coupling
is given by the laser intensity. 

\section*{D. Nuclear wave packet propagation and calculated quantities}

The non-adiabatic nuclear dynamics is governed by the Hamiltonian
in Eq. (1) which contains the vibrational and rotational degrees of
freedom of the molecule. To study the photodissociation in the LICI
framework, one has to solve the time-dependent Schrödinger equation
(TDSE) by applying this Hamiltonian. One of the most efficient approaches
for this is the MCTDH (multiconfigurational time-dependent Hartree)
method \cite{MCTDH1,MCTDH2,MCTDH3}. The vibrational degree of freedom
$(\mathrm{R})$ was characterized by using FFT-DVR (Fast Fourier Transformation-Discrete
Variable Representation) with $\mathrm{N_{R}=512}$ grid points distributed
within the range from $2.0\,\mathrm{au}$ to $50\,\mathrm{au}$ for
the internuclear separation. The rotational degree of freedom was
described by Legendre polynomials $\mathrm{\left\{ P_{J}(\cos\theta)\right\} _{j=0,1,2,\cdots,N_{\theta}}}$.
These so called primitive basis sets ($\chi$) were used to represent
the single particle functions ($\phi$), which in turn were applied
to represent the wave function: 
\begin{eqnarray}
\mathrm{\phi_{j_{q}}^{(q)}(q,t)} & = & \mathrm{\sum_{l=1}^{N_{q}}}\mathrm{c_{j_{q}l}^{(q)}(t)\;\chi_{l}^{(q)}(q)\qquad q=R,\,\theta}\label{eq:MCTDH-wf}\\
\mathrm{\psi(R,\mathrm{\theta,}t)} & = & \mathrm{\sum_{j_{R}=1}^{n_{R}}\sum_{j_{\theta}=1}^{n_{\theta}}}\mathrm{A_{j_{R},j_{\theta}}(t)\phi_{j_{R}}^{(R)}(R,t)\phi_{j_{\theta}}^{(\theta)}(\theta,t)}.\nonumber 
\end{eqnarray}
 In the actual calculations, we have used $\mathrm{N_{\theta}=91}$
and for both degrees of freedom a set of $\mathrm{n_{R}=n_{\theta}=28}$
single particle functions were applied to construct the nuclear wave
packet. In all simulations, the convergence was tested with these
parameters. 

The solution of the nuclear TDSE with the ansatz in equation \ref{eq:MCTDH-wf}
is utilized to calculate the kinetic energy release (KER) $\mathrm{P_{KER}(E)}$,
the total dissociation probability $\mathrm{P_{diss}}$ and the angular
distribution $\mathrm{P(\theta_{j})}$ of the dissociation fragments.
The kinetic energy release (KER) of the photofragments is \cite{MCTDH3}:

\begin{equation}
\mathrm{P_{KER}(E)}=\mathrm{\intop_{0}^{\infty}dt\intop_{0}^{\infty}dt'}\mathrm{<\psi(t)|W|\psi(t')>e^{-iE(t-t')}}\label{eq:ker}
\end{equation}
 where $\mathrm{-iW}$ is the complex absorbing potential (CAP) applied
at the last $\mathrm{10\,au}$ of the grid related to the vibrational
degree of freedoms ($\mathrm{W=0.000012\cdot\left(\mathrm{\mathrm{R}}-40\right)^{3}}$,
if $\mathrm{\mathrm{\mathrm{R}>40}\,au}$ on all three surfaces)  and

\begin{equation}
\mathrm{P(\theta_{j})}=\mathrm{\frac{1}{w_{j}}\intop_{0}^{\infty}dt<\psi(t)|W_{\theta_{j}}|\psi(t)>}\label{eq:angdist}
\end{equation}
 where $\mathrm{-iW_{\theta_{j}}}$ is the projection of the CAP to
a specific point of the angular grid $\mathrm{\left(j=0,..N_{\theta}\right)}$,
and $\mathrm{w_{j}}$ is the weight related to this grid point according
to the applied DVR. 

The total dissociation probability can be determined as \cite{MCTDH3}:

\begin{equation}
\mathrm{P_{diss}=\intop_{0}^{\infty}dE\cdot P_{KER}(E)}.\label{eq:tot-diss}
\end{equation}

Throughout the calculations, the initial nuclear wave function (at
$\mathrm{t\ll\mathrm{t_{pm}}}$) was assumed to be in its vibrational
and rotational ground state ($\mathrm{\nu=0;\:J=0})$. Initially,
the molecules are assumed in the numerical calculations to be nonaligned
and isotropically distributed and subject to linearly polarized $\cos^{2}$
pump and probe laser pulses centered around $\mathrm{t_{pm}=0\,fs}$
and $\mathrm{t_{pr}=0-1000\,fs}$, respectively. 

To demonstrate the impact of light-induced non-adiabatic effect on
the dissociation dynamics of the $\mathrm{\mathrm{NaH}}$, results
obtained from the full two-dimensional (2D) and one-dimensional (1D)
models were compared. In the 2D case both the rotational and vibrational
coordinates are accounted for as dynamical variables while in the
1D situation, the rotational degree of freedom ($\mathrm{\theta}$)
is used only as a parameter and accordingly, only LIACs and not LICI
can be considered. In the 1D model the molecule's initial orientation
is not changing during the dynamics and the ``effective field strength''
was the projection of the real field on the direction of the actual
dipole moment. Both the transition dipole between the $\mathrm{V_{X}(R)}$
and $\mathrm{V_{A}(R)}$ energy states and all the permanent dipoles
are parallel to the axis of the molecule, so $\mathrm{\varepsilon_{0}^{eff}(\theta)=\varepsilon_{0}\cos\theta}$
(i.e., with an effective intensity of $\mathrm{I_{0}^{eff}(\theta)=I_{0}\cos^{2}\theta}$).
For the transition dipoles related to a $\Sigma\,\longleftrightarrow\,\Pi$
electronic transition ($\mu_{XB}$ and $\mu_{AB}$) the dipoles are
perpendicular to the molecular axis, therefore $\mathrm{\varepsilon_{0}^{eff}(\theta)=\varepsilon_{0}\sin\theta}$
(i.e., with an effective intensity of $\mathrm{I_{0}^{eff}(\theta)=I_{0}\sin^{2}\theta}$). 

\part*{III. Results and discussion }

In this section, we investigate the non-adiabatic dissociation dynamics
of a molecular system. As mentioned in the introduction our aim is
to perform a pump and probe numerical simulation for the $\mathrm{NaH}$
molecule, so as to demonstrate the impact of the light-induced non-adiabatic
phenomena on the dissociation dynamics. To study this effect the three
lowest-lying singlet electronic states and an initial isotropic distribution
of the molecular ensemble are considered. The corresponding transition
dipole matrix elements are responsible for the light-induced electronic
transitions. 

In Figure 2 the corresponding potential energy surfaces and the light-dressed
states are visualized. At first, by using a pump pulse a part of the
ground $\mathrm{V_{X}(R)}$ electronic state population is transferred
to the first excited $\mathrm{V_{A}(R)}$ state. Then with a time
delay the probe pulse couples the$\mathrm{V_{A}(R)}$ state to the
$\mathrm{V_{X}(R)}$ and $\mathrm{V_{B}(R)}$ ones by transferring
further a reasonable amount of the $\mathrm{V_{A}(R)}$ population
to these states. With the pump pulse applied, more than 50\% of the
ground state population is transferred from the $\mathrm{V_{X}(R)}$
state to the $\mathrm{V_{A}(R)}$ one. The dynamics therefore is initiated
on the $\mathrm{V_{A}(R)}$ state, but the essential part of the dissociation
occurs on the $\mathrm{V_{X}(R)}$ and $\mathrm{V_{B}(R)}$ states.
The $\mathrm{V_{B}(R)}$ is already a dissociative state, but dissociation
can also occur on $\mathrm{V_{X}(R)}$ due to the fact that the population
arrives here with sufficiently large kinetic energy. 

\begin{figure}
\begin{centering}
\includegraphics[width=0.48\textwidth]{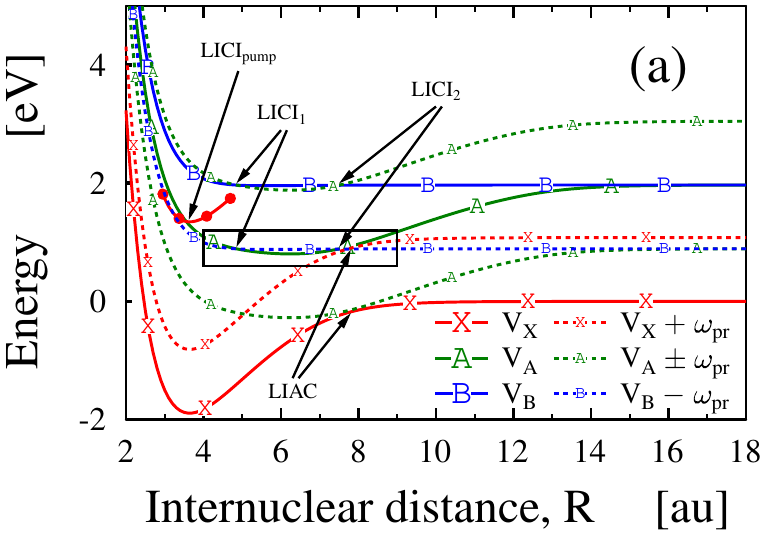}\includegraphics[width=0.48\textwidth]{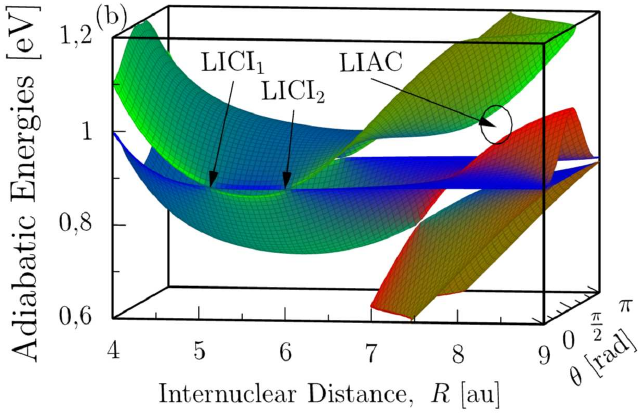}
\par\end{centering}
\caption{\label{fig:fig2} (a)The three lowest-lying singlet adiabatic potential
energy curves of the $\mathrm{NaH}$ molecule. The corresponding light-dressed
states are denoted by dashed lines. The position of the light-induced
conical intersections (LICI$\mathrm{_{pump}},$ LICI1 and LICI2) and
avoided crossing (LIAC) are also marked. (b)Light-induced potential
energy surfaces for the three-state model calculated for the field
intensity $\mathrm{I_{pr}=}$$1\cdot10^{12}\mathrm{W/cm^{2}}$. The
coloring reflects the diabatic states that make up the different points
of the adiabatic surfaces: red, $\mathrm{V_{X}}(\mathrm{R)}$; green,$\mathrm{V_{A}}(\mathrm{R)}$;
and blue,$\mathrm{V_{B}}(\mathrm{R)}$.}
\end{figure}

Figure 3 shows the field-free time evolution of the nuclear wave packet
on the $\mathrm{V_{A}}(\mathrm{R})$ potential energy surface after
using a pump pulse. The field-free wave packet oscillates back and
forth on the interval of $\mathrm{R\approx3-11\,\mathrm{au}}$. Initially
and in the first quarter of the studied time domain it is relatively
well localized, while in the last quarter it is rather diffuse. Although
it is still possible to recognize its periodicity in this latter time
domain, but it can no longer be localized with a relatively good approximation.
The dashed lines (at $\mathrm{R=4.85}$ au and $\mathrm{R=7.5}$ au)
are the positions of the LICI1 and LICI2 between the $\mathrm{V_{A}}(\mathrm{R})$
and $\mathrm{V_{B}}(\mathrm{R})$ surfaces, while the dotted line
(at $\mathrm{R=7.8}$ au) is the position of the LIAC between the
$\mathrm{V_{A}}(\mathrm{R})$ and $\mathrm{V_{X}}(\mathrm{R})$. It
is expected that the dissociation yields on the corresponding surfaces
will be large for those values of the $\mathrm{t_{del}}$ where these
horizontal lines intersect with the nuclear wave packet (at the position
of LICI1, LICI2 and LIAC). Namely, at $\mathrm{\mathrm{t}\approx15,90,110...fs}$
on the $\mathrm{V_{B}}(\mathrm{R})$ state and at $\mathrm{\mathrm{t}\approx30,70,..fs}$
on the $\mathrm{V_{X}}(\mathrm{R})$ state. We note here that due
to the large transition dipole moment values (see Fig. 1b), the influence
of LICI1 and LIAC on the non-adiabatic dynamics is more significant
than that of the LICI2. 

\begin{figure}
\begin{centering}
\includegraphics[width=0.48\textwidth]{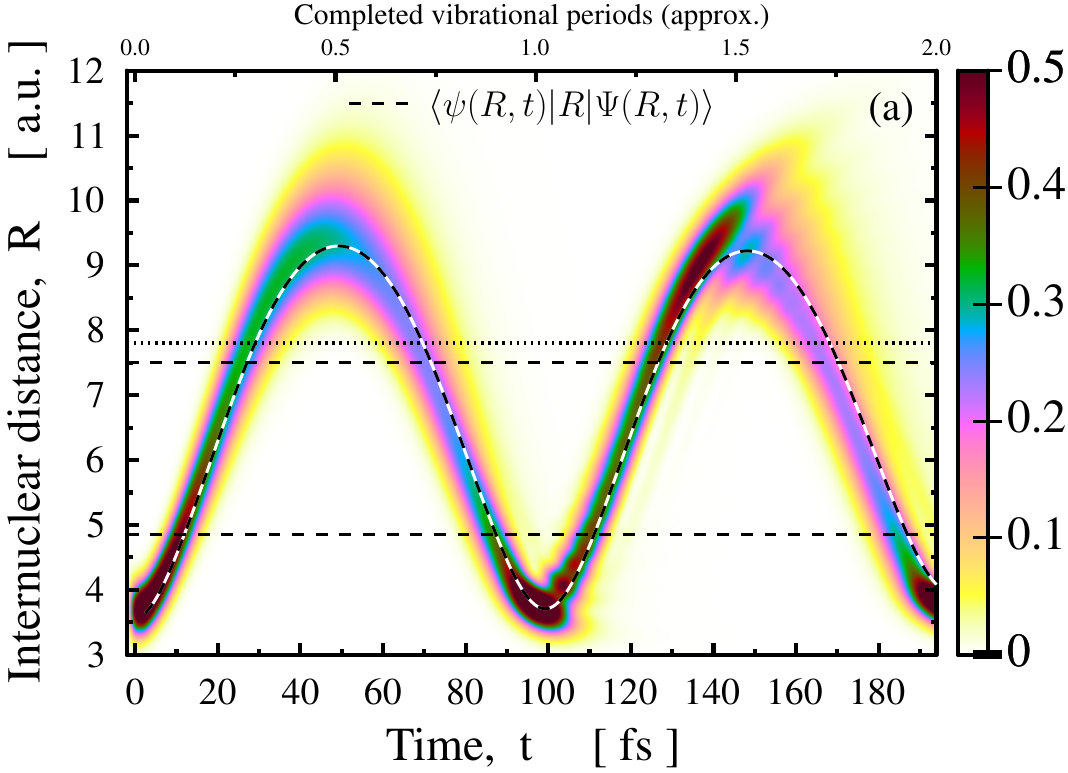}\includegraphics[width=0.48\textwidth]{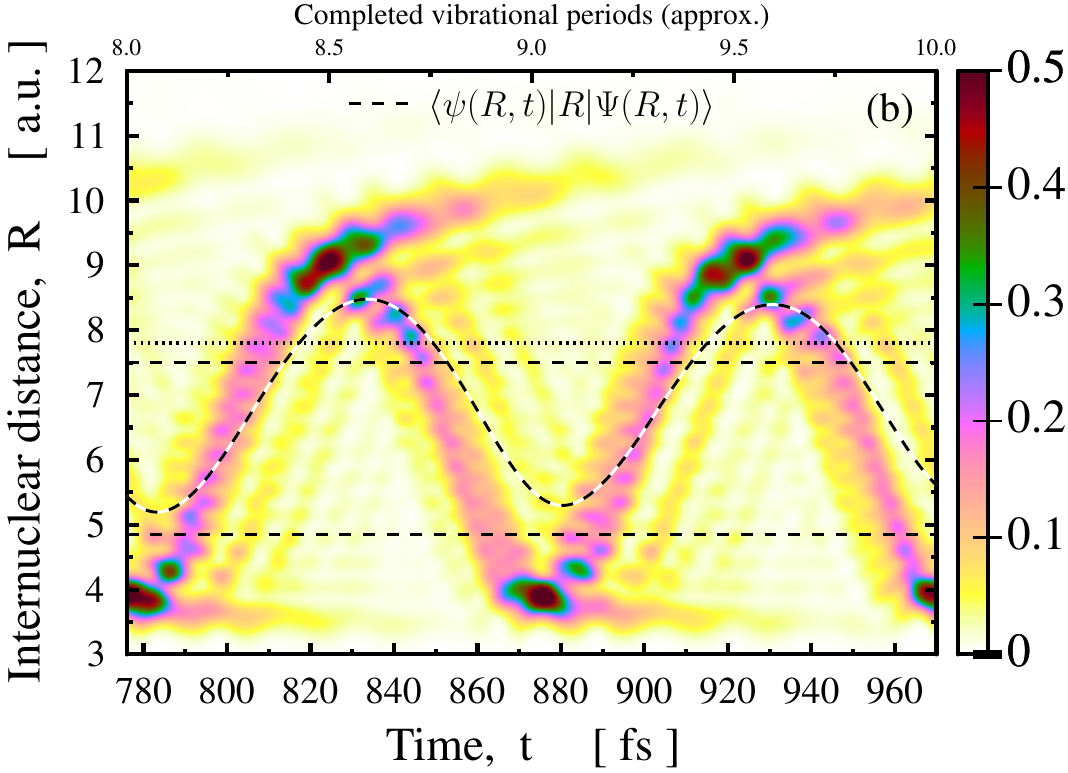}
\par\end{centering}
\caption{\label{fig:fig3} Field-free vibrational motion of the nuclear wave
packet -- after applying the pump pulse -- on the $\mathrm{V_{A}}(\mathrm{R})$
potential energy curve. The time dependent density of the nuclear
wave packet is displayed in an initial ($\mathrm{t\approx0-190\,\mathrm{fs}}$)
and a final ($\mathrm{t\approx750-980\,fs}$) time periods by color
map. The darker the color, the larger the value of the nuclear density.
In addition, the time-dependent average of the internuclear distance
(dashed line) is also shown. The dashed lines (at $\mathrm{R=4.85}\,\mathrm{au}$
and $\mathrm{R=7.5}\,\mathrm{au}$) are the positions of the LICI1
and LICI2 between $\mathrm{V_{A}}(\mathrm{R})$ and $\mathrm{V_{B}}(\mathrm{R})$,
while the dotted line (at $\mathrm{R=7.8\,\mathrm{au}}$) is the position
of the LIAC between $\mathrm{V_{A}}(\mathrm{R})$ and $\mathrm{V_{X}}(\mathrm{R})$. }
\end{figure}

In Fig. 4 the dissociation probabilities correspond to each of the
individual states, as well as the sum of them as a function of time
delay ($\mathrm{t_{del}}$) are displayed. At first glance, it is
obvious that a significant part of the dissociation takes place from
surface $\mathrm{V_{B}}(\mathrm{R}$). The dissociation probability
of the $\mathrm{V_{X}(R)}$ state at certain time delays can be up
to 10-12\%, while in the case of $\mathrm{V_{B}(R)}$ it can even
go over 40\% of the whole population. The dissociated population of
the $\mathrm{V_{X}(R)}$ state originates from the population of the
initially populated $\mathrm{V_{A}(R)}$ state. For the case of the
$\mathrm{V_{B}(R)}$ state the 2\% of the dissociated population originates
from the effect of the pump pulse, and the remaining large amount
of the dissociated population arises due to the population transfer
of the $\mathrm{V_{A}(R)}$ state by the probe pulse, as for the case
of the $\mathrm{V_{X}(R)}$. From the third state $\mathrm{V_{A}(R)}$,
the dissociation yield is about two orders of magnitude smaller and
therefore it is practically invisible on the scale used. Consequently,
the period of the total dissociation yield is practically dominated
by the period of the dissociation yield occurring on the $\mathrm{V_{B}}(\mathrm{R}$).
One can also notice that the structure of the dissociation yields
obtained from the $\mathrm{V_{X}}(\mathrm{R}$) and $\mathrm{V_{B}}(\mathrm{R}$)
states, as well as the curve of the total yield are different. The
origin of the difference can easily be understood based on Fig. 3.
The dissociation yields possess large values at those $\mathrm{t_{del}}$
delay times where the field-free wave packet has a significant amplitude
in the close vicinity of the LICIs and LIAC positions. These points
are the intersection points of the wave packet and the horizontal
lines in Fig. 3. The intersection points -- between the position
of LICI1 and the wave packet -- are closer to the turning point of
the nuclear vibration, therefore the peaks on the yield curve from
the $\mathrm{V_{B}}(\mathrm{R})$ state are also closer to each other
than that of the $\mathrm{V_{X}}(\mathrm{R})$ state. The two panels
of Fig. 4 compare the results of the 1D and 2D calculations. It can
be seen that the amplitude of the dissociation yields decrease monotonically
as a function of time delay, and this decreasing is more prominent
in the 2D results. Although the magnitude of the amplitudes decreases,
their minimum values continuously increase, while their maximum values
continuously decrease as a function of time delay. This effect is
due to the increasing blurring of the nuclear wave packet at increasing
$\mathrm{t_{del}}$ values. The differences between the 1D and 2D
results can be interpreted, as a fingerprint of the molecular rotation
combined with the effects of the LICIs and LIAC as well as with the
different magnitudes/directions of the respective transition dipole
moments. The pump pulse transfers a part of the population to surface
$\mathrm{V_{A}}(\mathrm{R})$, and at the same time initiates the
rotation of the wave packet. After the pump process, as time increases,
the molecular axis rotates closer and closer to the polarization direction
of the pump pulse. On the one hand, when the probe appears, it transfers
further a part of the population from $\mathrm{V_{A}}(\mathrm{R}$)
to $\mathrm{V_{B}}(\mathrm{R}$). Because the transition dipole moment
between these two states is perpendicular to the molecular axis, it
can transfer less population than in the case of 1D, when the rotation
is frozen. On the other hand, the probe can also couple the $\mathrm{V_{A}}(\mathrm{R})$
and $\mathrm{V_{X}}(\mathrm{R})$ states and it provides more efficient
population transfer due to the parallel direction of the $\mathrm{\mu_{A,X}}$
to the polarization direction of the probe pulse. Consequently, it
starts to transfer back some parts of the population to the $\mathrm{V_{A}}(\mathrm{R})$
state and Rabi oscillation starts between these two states. Finally,
it slightly suppresses the total dissociation yield relative to the
1D model.

\begin{figure}
\begin{centering}
\includegraphics[width=0.48\textwidth]{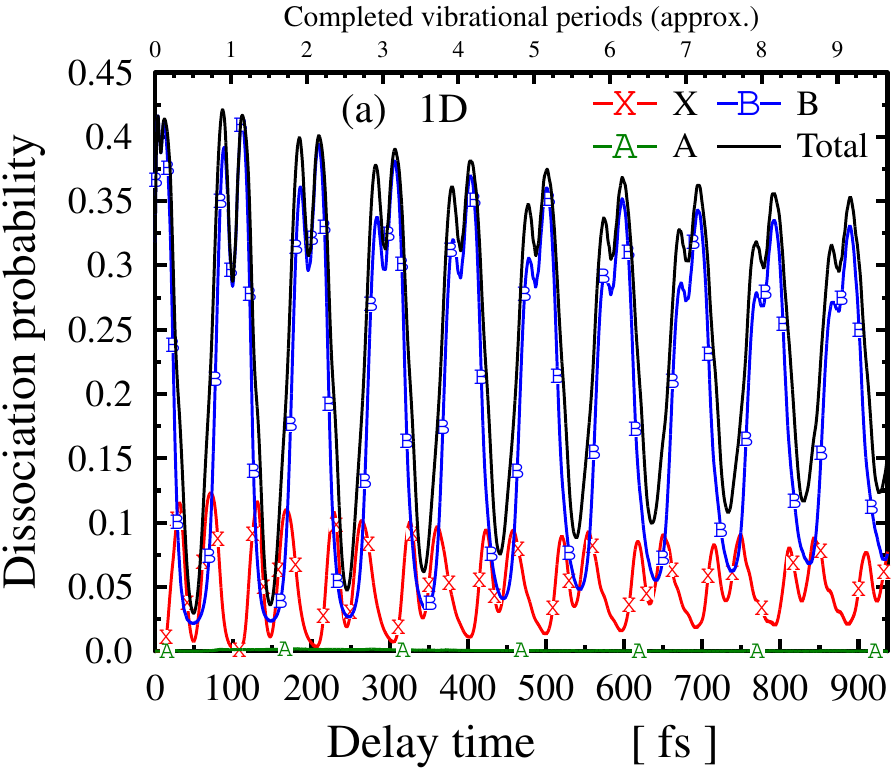}\includegraphics[width=0.48\textwidth]{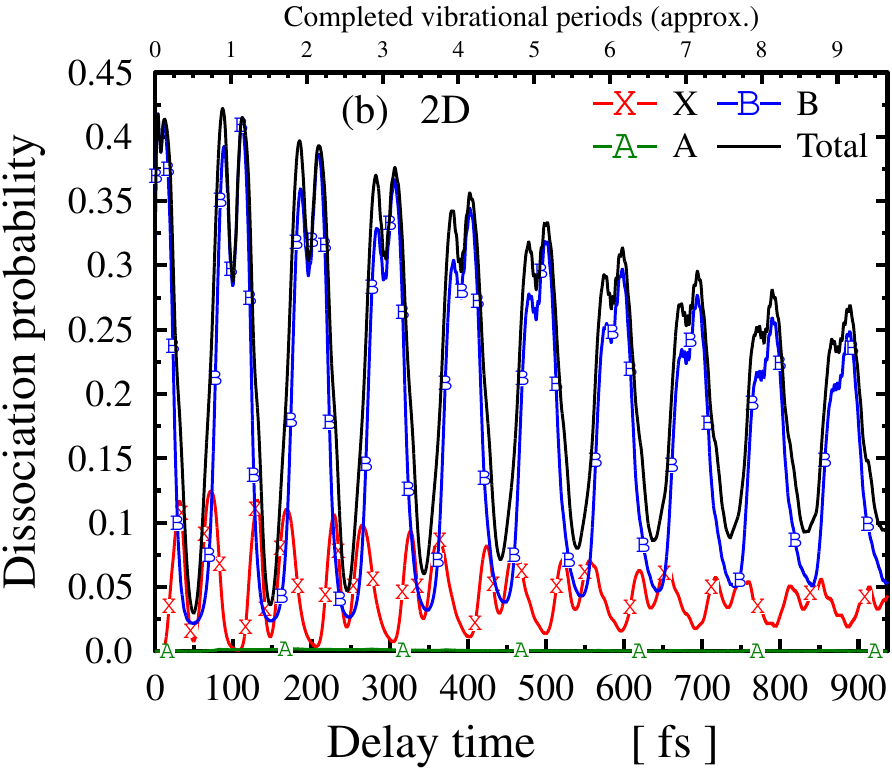}
\par\end{centering}
\caption{\label{fig:fig4} Dissociation probabilities corresponding to each
of the individual adiabatic electronic states as well as the sum of
them as a function of time delay between the pump and probe pulses.
Calculations are displayed both in the 1D and 2D models (see on panel
(a) and (b), respectively). }
\end{figure}
\begin{figure}
\begin{centering}
\includegraphics[width=0.33\textwidth]{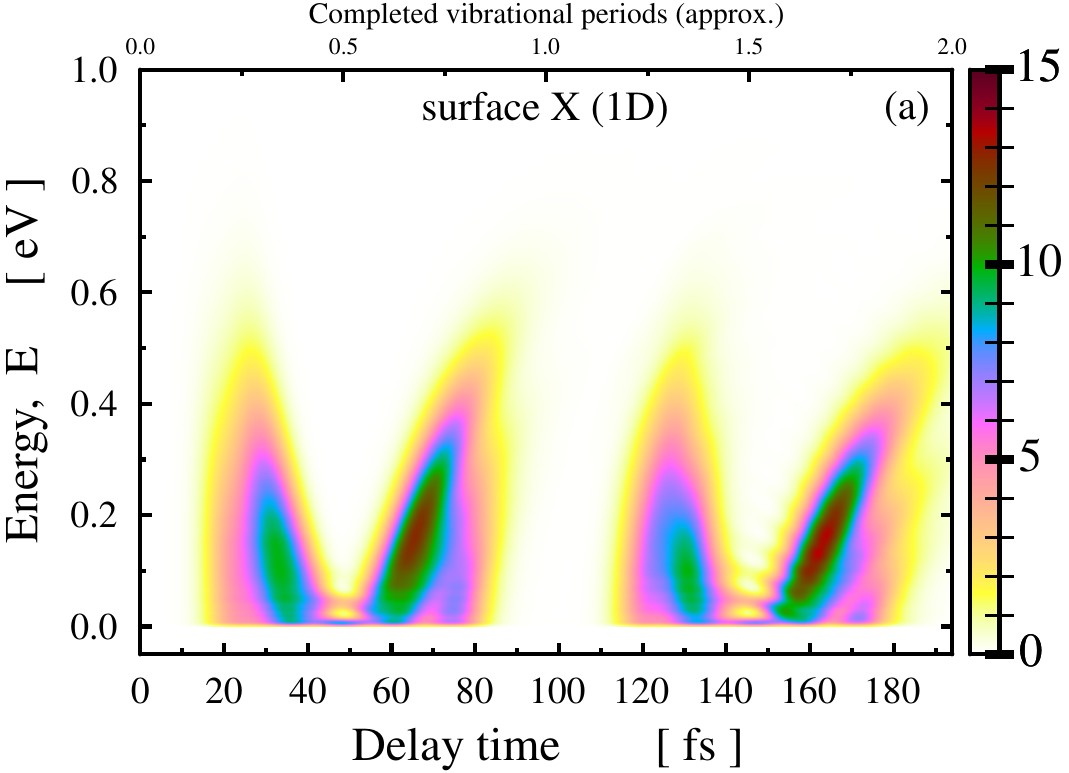}\includegraphics[width=0.33\textwidth]{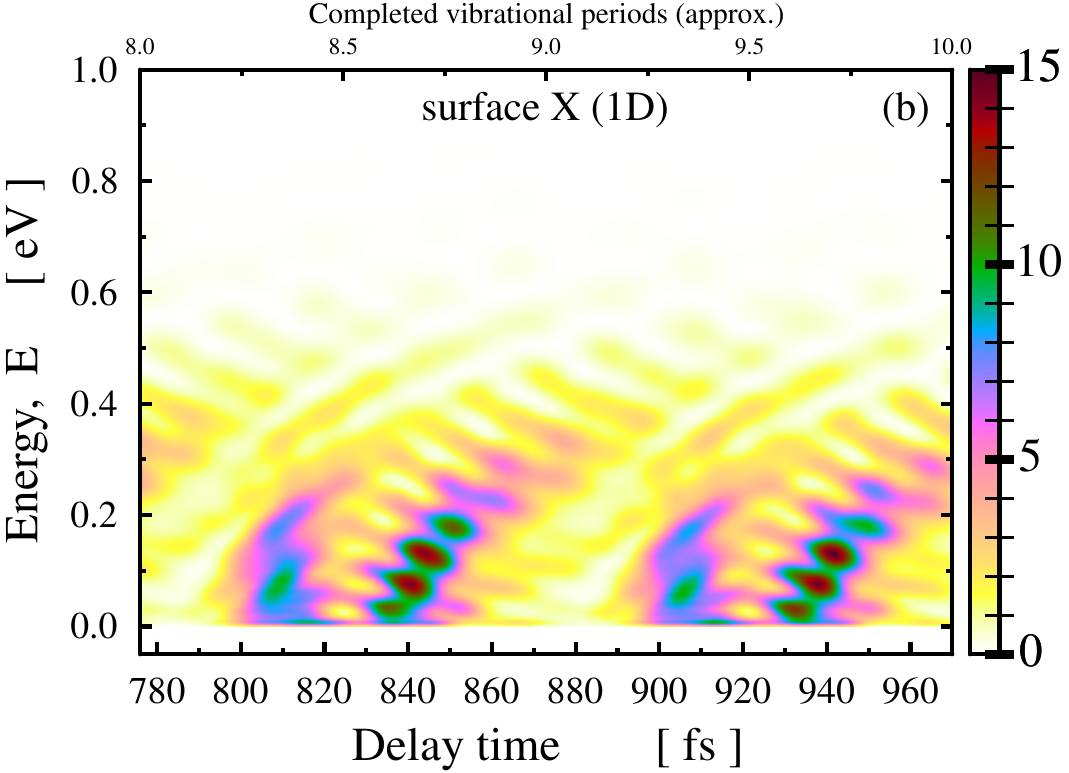}\includegraphics[width=0.33\textwidth]{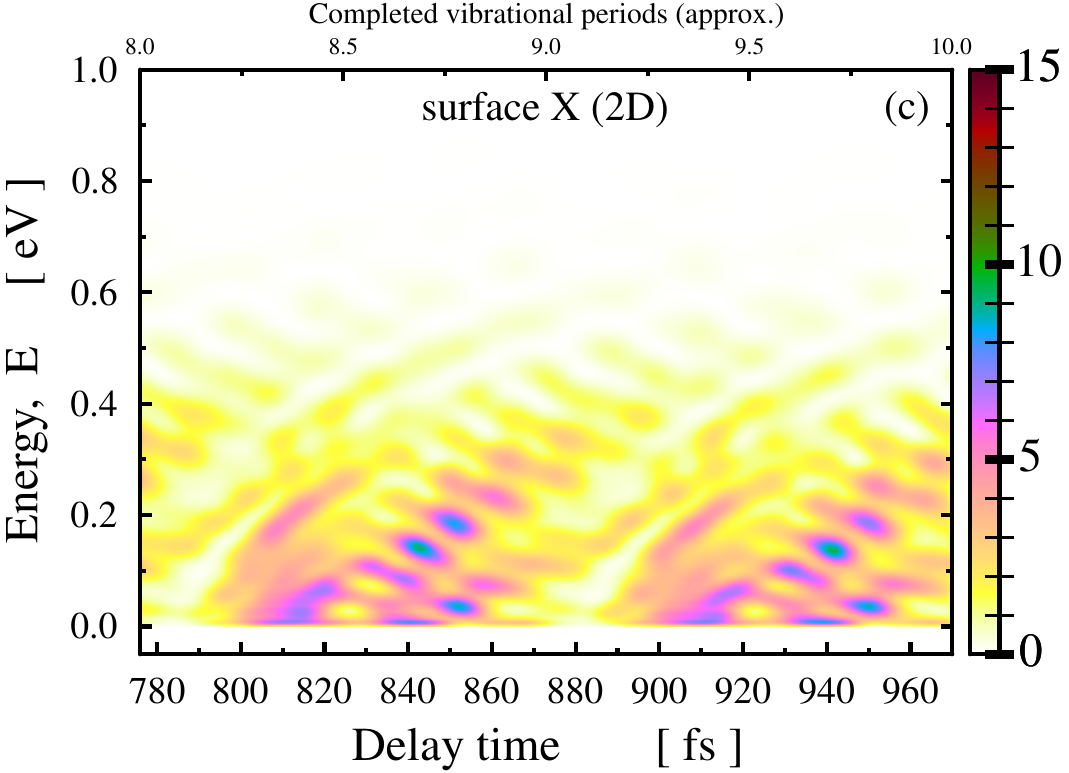}
\par\end{centering}
\begin{centering}
\includegraphics[width=0.33\textwidth]{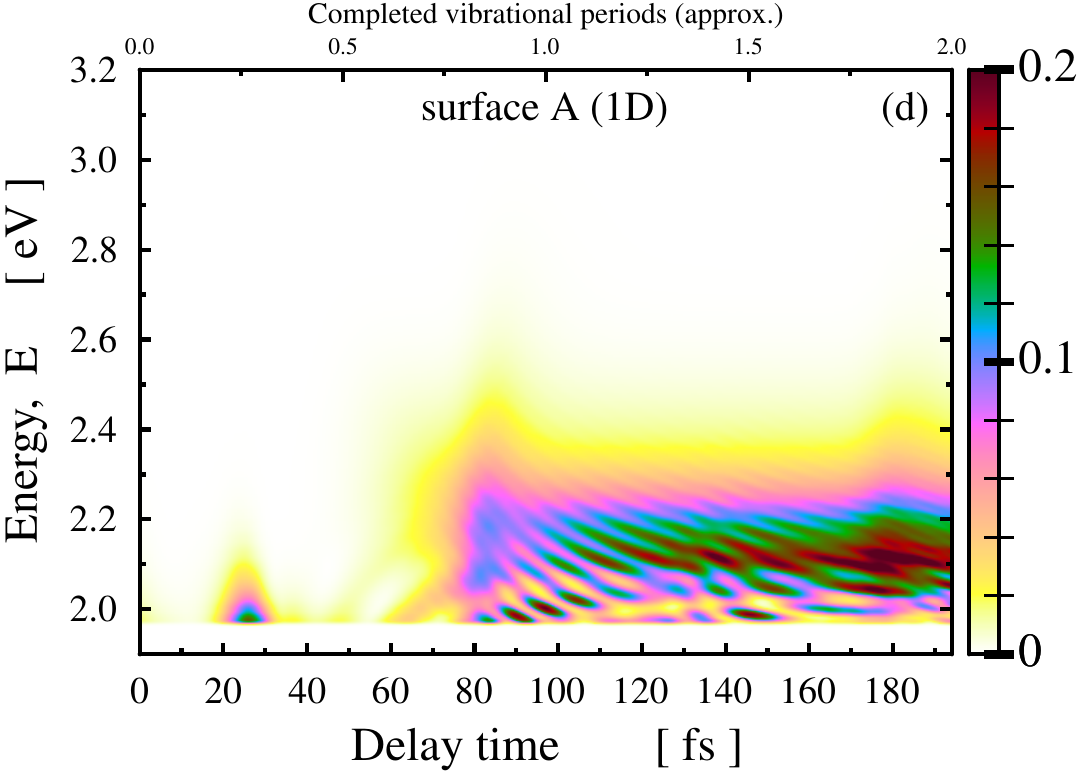}\includegraphics[width=0.33\textwidth]{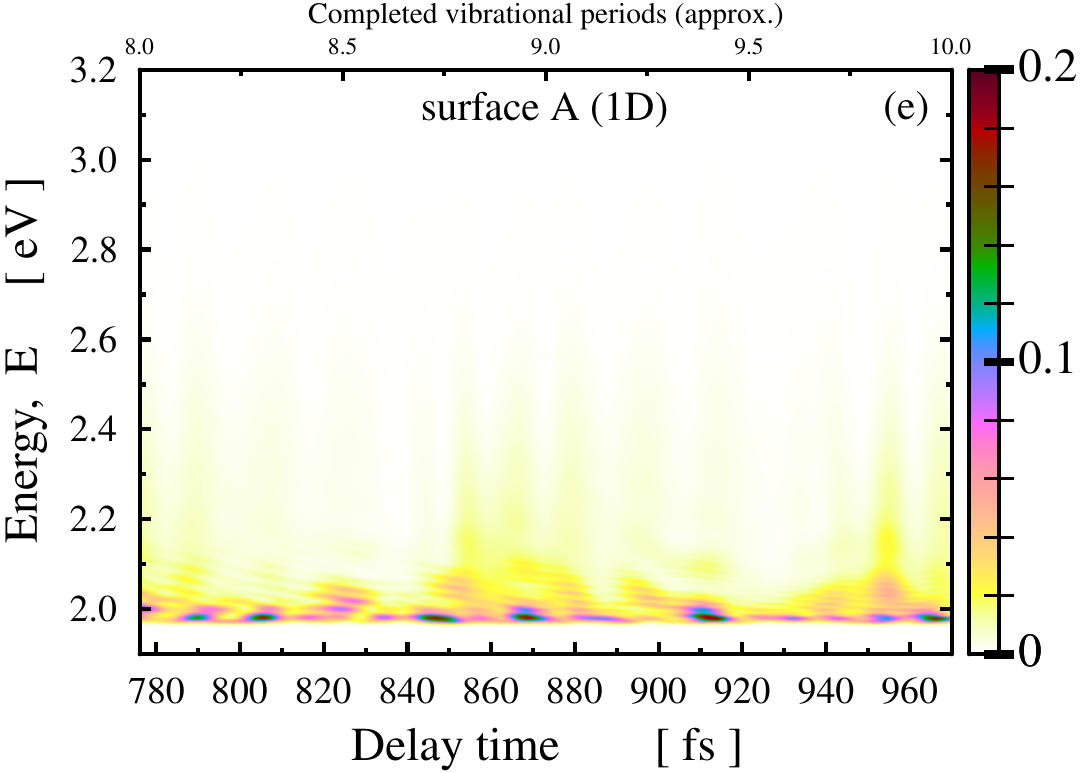}\includegraphics[width=0.33\textwidth]{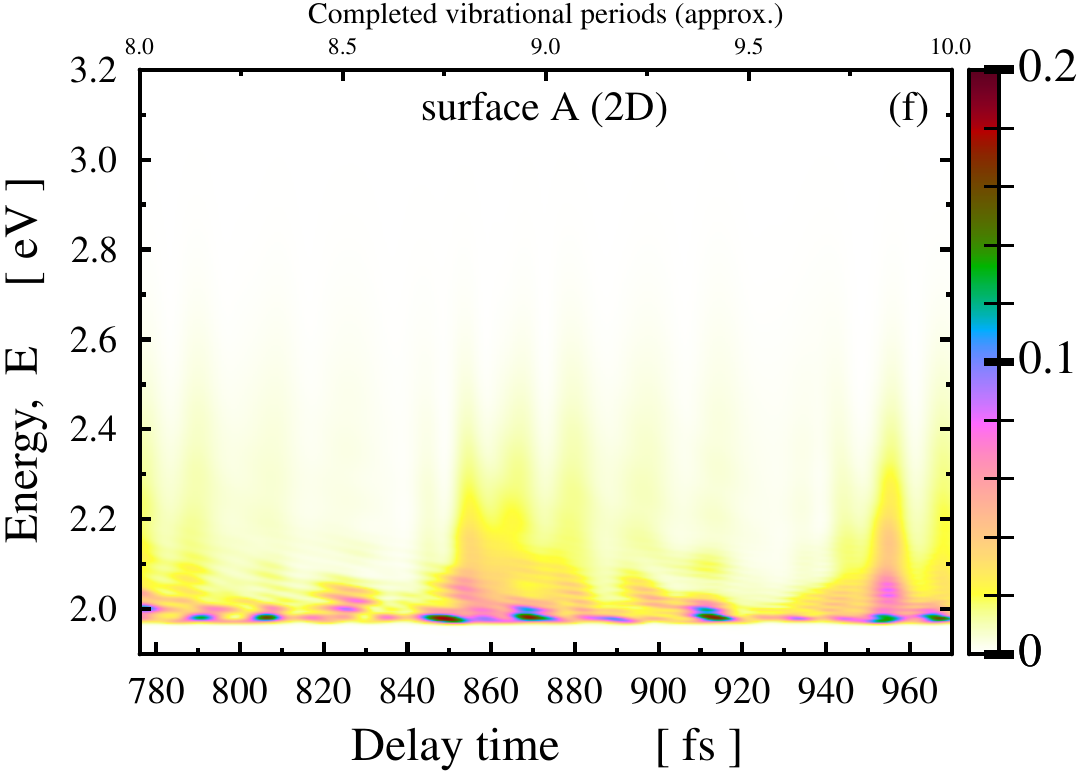}
\par\end{centering}
\begin{centering}
\includegraphics[width=0.33\textwidth]{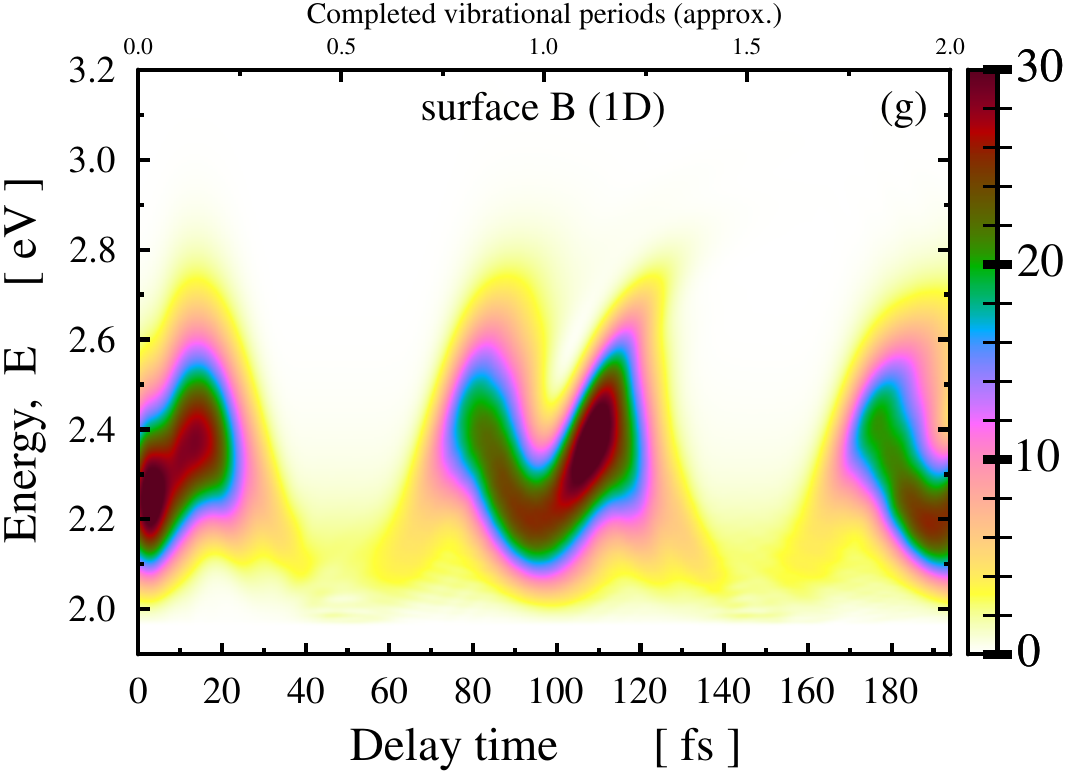}\includegraphics[width=0.33\textwidth]{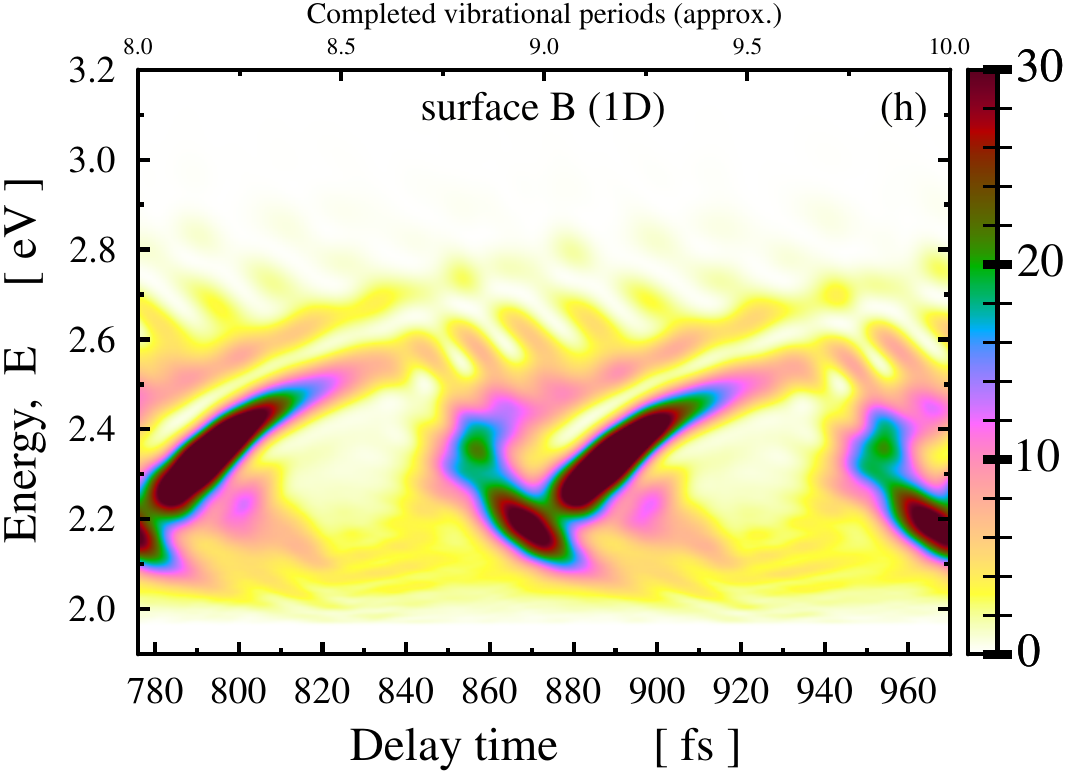}\includegraphics[width=0.33\textwidth]{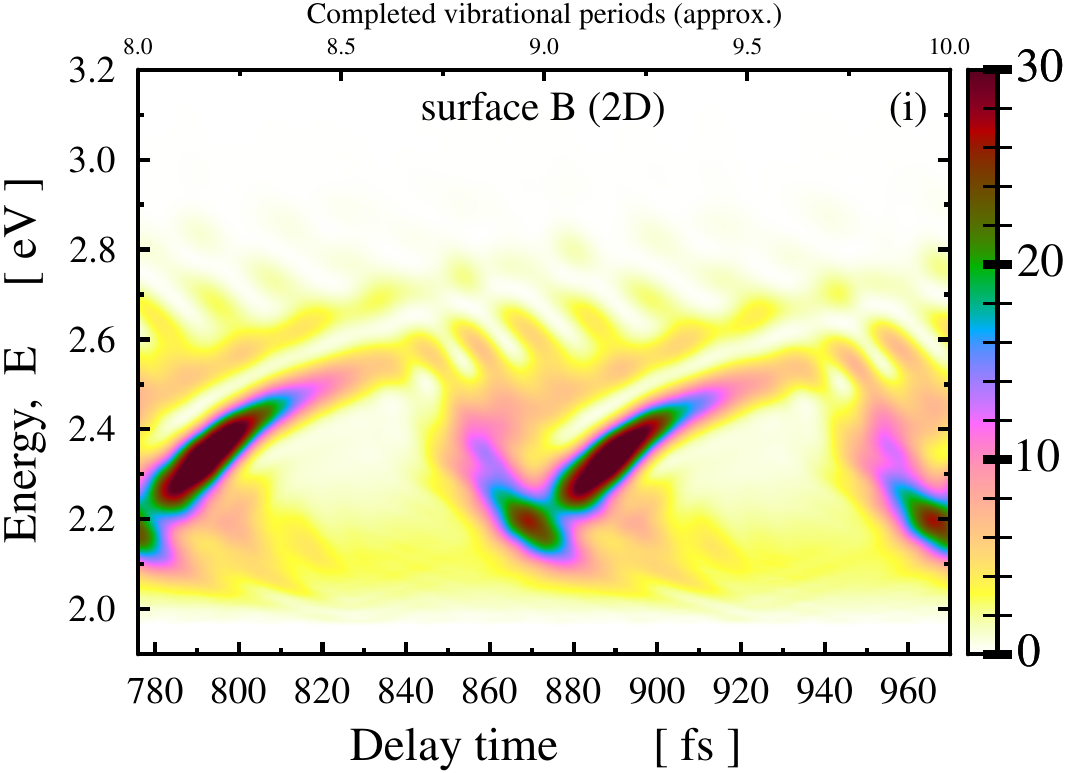}
\par\end{centering}
\caption{\label{fig:fig5} Kinetic energy release spectra (KER) of the photofragment
of the $\mathrm{NaH}$ molecule as a function of delay time. Results
are presented for each of the individual channels in the 1D as well
as in the 2D framework. }
\end{figure}

To clearly visualize the dependence of the kinetic energy release
(KER) of the photofragments as a function of time delay, results for
the three individual states are displayed in Fig. 5. For the case
of shorter delay times the 1D and 2D results are quite the same, therefore
only the 1D spectra are presented. The fact that the two different
models are almost identical for these short delay times means that
the rotation initialized by the pump pulse on surface $\mathrm{V_{A}}(\mathrm{R})$
has not yet been significantly developed before the probe pulse hits
the system. Contrary to this, for larger delay times, the difference
between the two descriptions is already noticeable. Here, in the photofragment
KER spectra of the different states, one can clearly recognize the
three-dimensional fingerprint of the dissociation yields obtained
from the corresponding channels. Namely, the 1D model always provides
a more significant dissociation yield on the $\mathrm{V_{X}}(\mathrm{R})$
and $\mathrm{V_{B}}(\mathrm{R})$ states than that of the 2D one.
\begin{figure}
\begin{centering}
\includegraphics[width=0.48\textwidth]{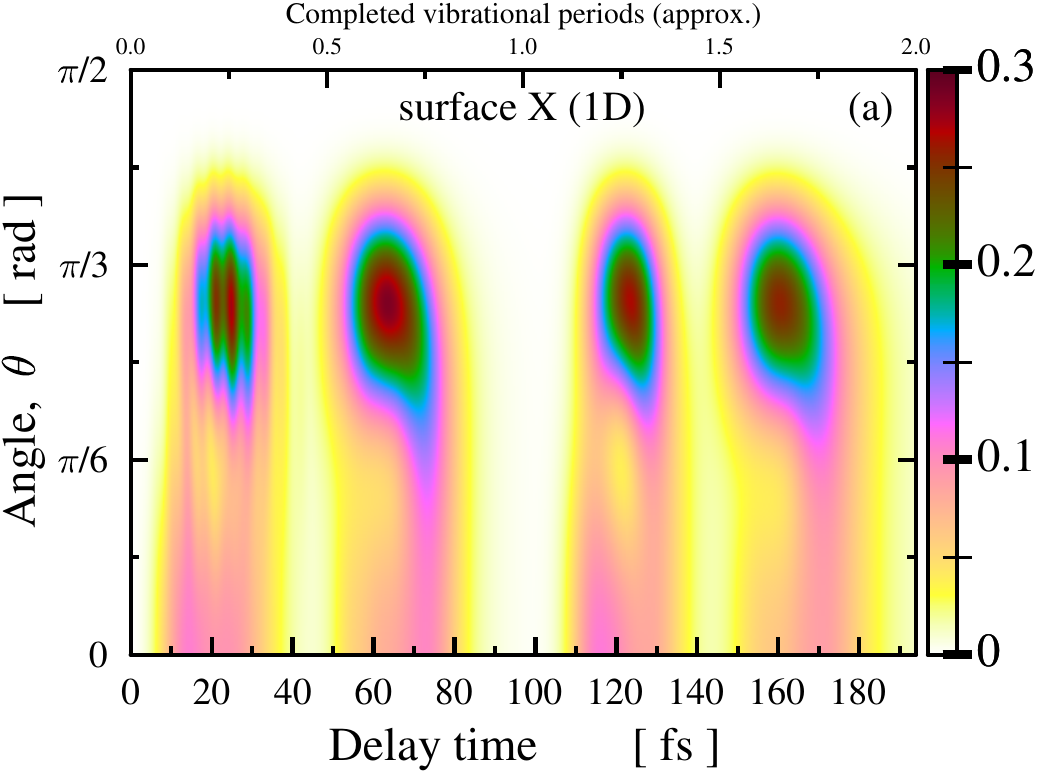}\includegraphics[width=0.48\textwidth]{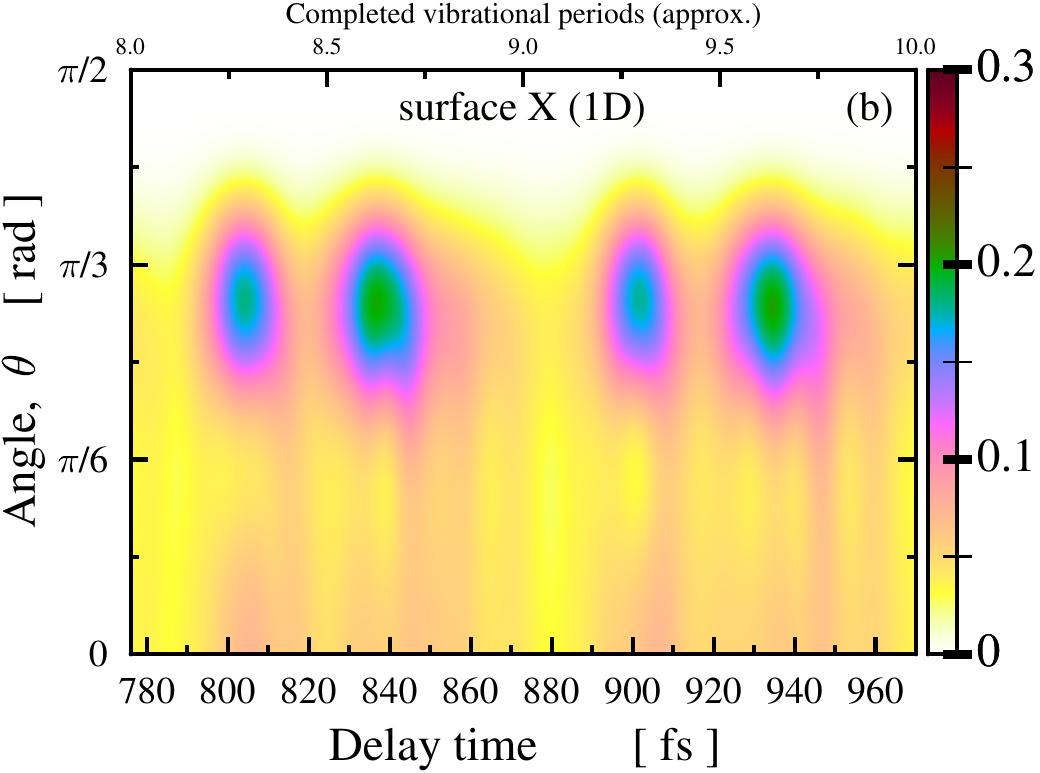}
\par\end{centering}
\begin{centering}
\includegraphics[width=0.48\textwidth]{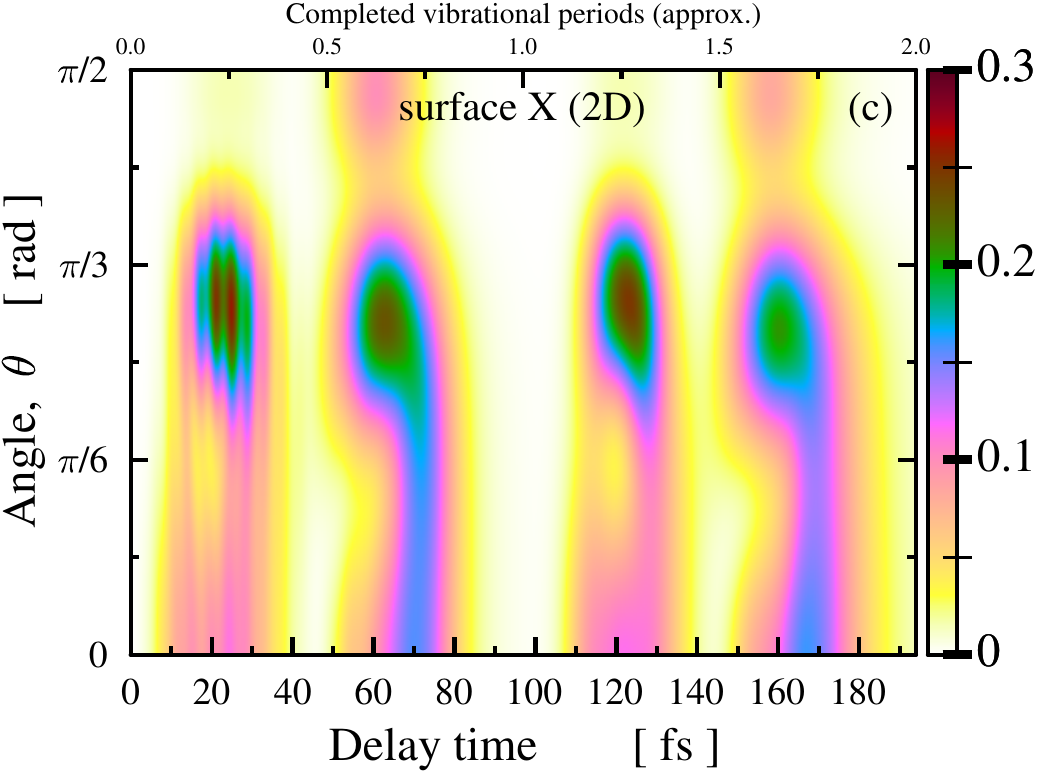}\includegraphics[width=0.48\textwidth]{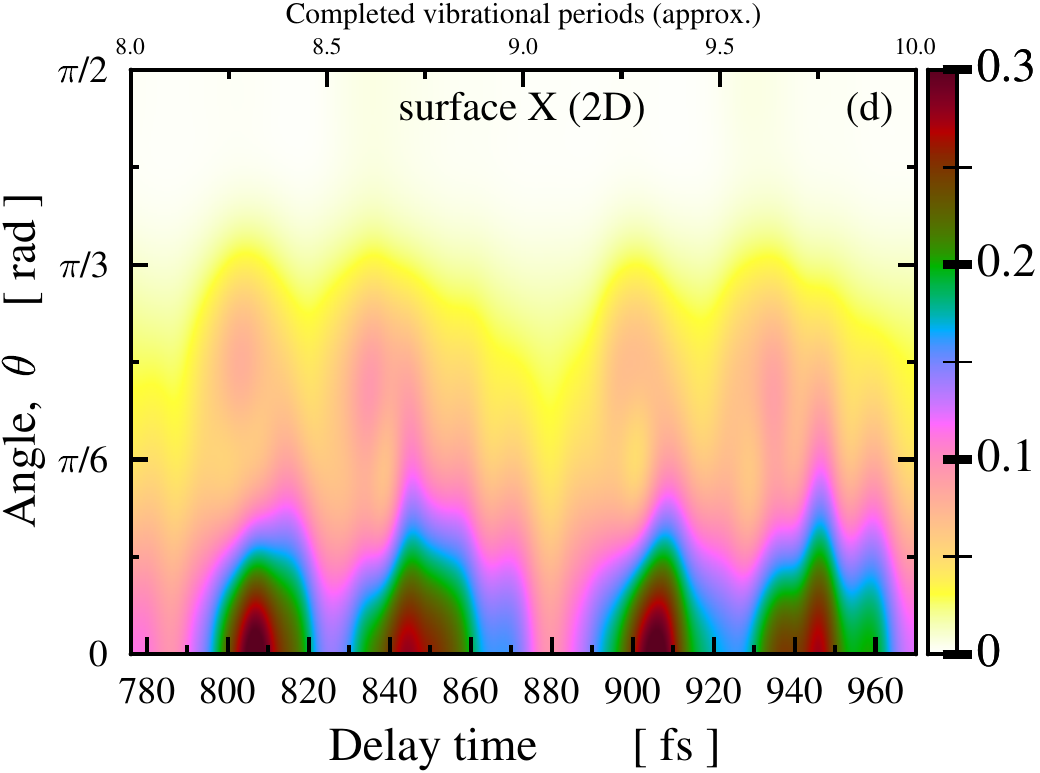}
\par\end{centering}
\caption{\label{fig:fig6}Fragment angular distribution of the dissociating
NaH molecule for the $\mathrm{V_{X}}(\mathrm{R}$) electronic state
as a function of time delay. Results are presented both for the 1D
and 2D schemes.}
\end{figure}

To further analyze the dissociation process, the fragment angular
distributions of the $\mathrm{V_{X}}(\mathrm{R}$) ground electronic
state are displayed in Fig. 6. In panel (a) the 1D results are presented
for short delay times. In this case, the molecules do not rotate and
because the transition dipole moment is parallel to the molecular
axis between the $\mathrm{V_{X}}(\mathrm{R})$ and $\mathrm{V_{A}}(\mathrm{R}$)
states, nothing appears at around $\mathrm{\mathrm{\theta=90}}$ degrees.
The pump pulse does not only transfer the population to the $\mathrm{V_{A}}(\mathrm{R}$),
but also takes it back to $\mathrm{V_{X}}(\mathrm{R}$). Similarly,
the probe does not only transfer from the $\mathrm{V_{A}}(\mathrm{R}$)
to $\mathrm{V_{X}}(\mathrm{R}$), but also takes it back to $\mathrm{V_{A}}(\mathrm{R}$).
Due to the larger coupling at around $\mathrm{\theta=0}$ degree,
two-photon process is also present. The periodic change of the maximum
values is consistent with the movement of the nuclear wave packet
(see Fig. 3). At larger delay times the fragment angular distribution
is much more spread out which is related to the diffuse character
of the nuclear wave packet (see Fig. 3b). However, the maximum values
remained at the same position as in the short $\mathrm{t_{del}}.$
Since the molecules cannot rotate, the positions of the maximum values
depend on what kind of molecular distribution has been developed right
immediately after the effect of the pump pulse. In Fig. 6c results
of the 2D simulations are presented for short delay times. The maximum
values of the 2D angular distributions are at the same position as
for the 1D case, but the effect of the rotation appears, as well.
Consequently, enhanced 2D dissociation occurs at certain time delays
($\mathrm{t_{del}=60\,\mathrm{fs}}$ and $\mathrm{t_{del}=160\,fs}$)
at $\mathrm{\theta=0}$ degree. Within this time interval, the difference
between the 1D and 2D schemes is not really significant, except for
that at $\mathrm{t_{del}=60\,\mathrm{fs}}$ and $\mathrm{t_{del}=160\,fs}$
time delays dissociation can take place at around $\theta=90$ degree.
The latter can be explained by the fact that the probe pulse rotates
the molecule to the direction of $\mathrm{\theta=0}$ degree on the
lower and to the direction of $\mathrm{\theta=90}$ degrees on the
upper adiabatic surfaces. Depending on in which state it was for longer
times, rotates towards $\mathrm{\theta=0}$ or $\mathrm{\mathrm{\mathrm{\theta=90}}}$
degrees during the dissociation. In contrast, at longer delay times
the situation is completely different (see Fig. 6d). Here, because
of the effect of the pump pulse the system has already had enough
time to rotate towards the polarization direction of the pump pulse
and consequently, the largest dissociation yield can be obtained at
around $\theta=0$ degree. The periodic structure is also visible
here, but it is much more blurred than at shorter delay times (see
Fig. 6c). 

\begin{figure}
\begin{centering}
\includegraphics[width=0.48\textwidth]{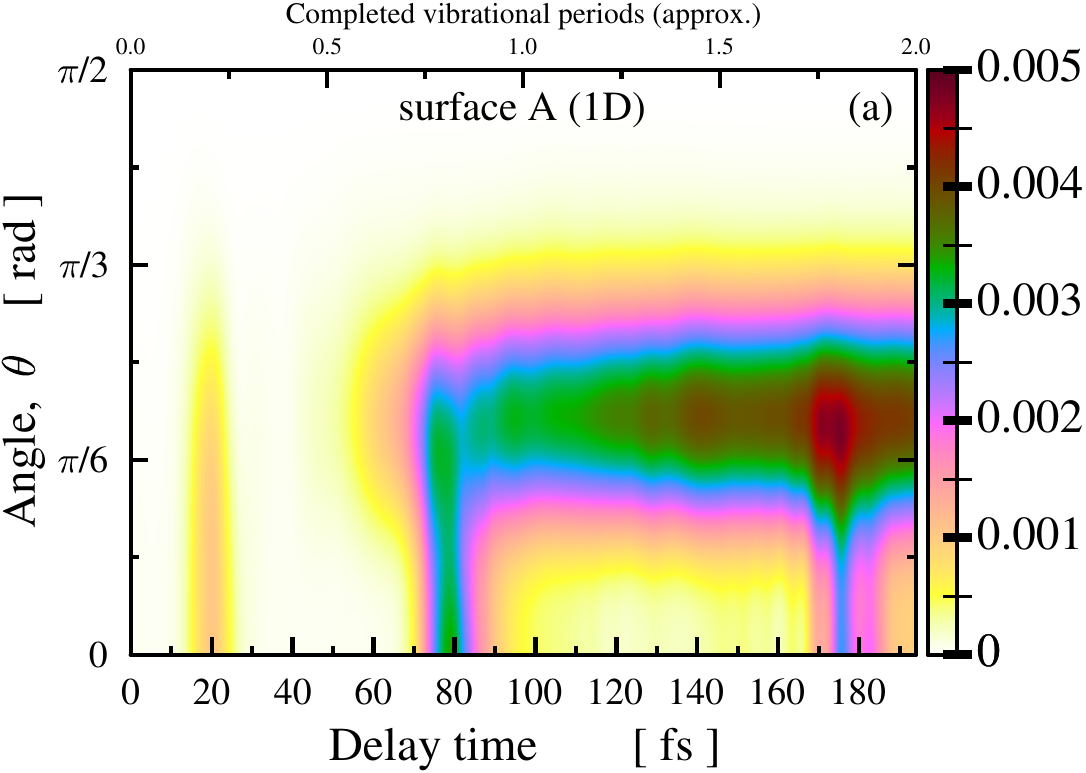}\includegraphics[width=0.48\textwidth]{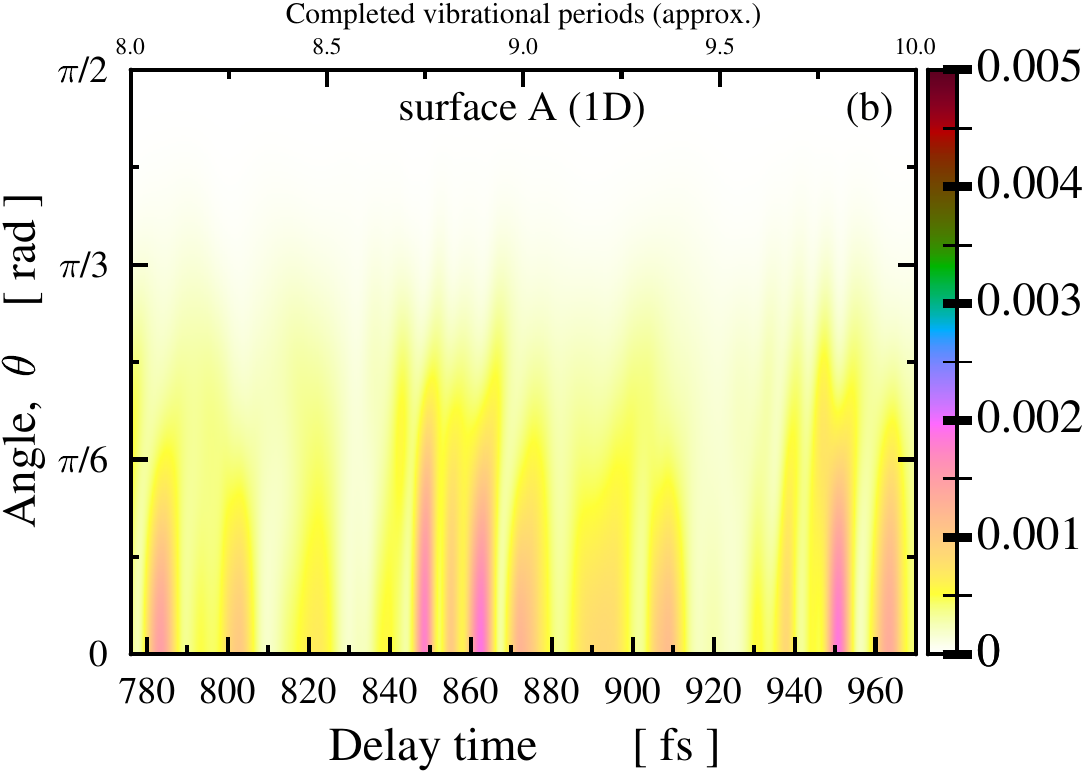}
\par\end{centering}
\begin{centering}
\includegraphics[width=0.48\textwidth]{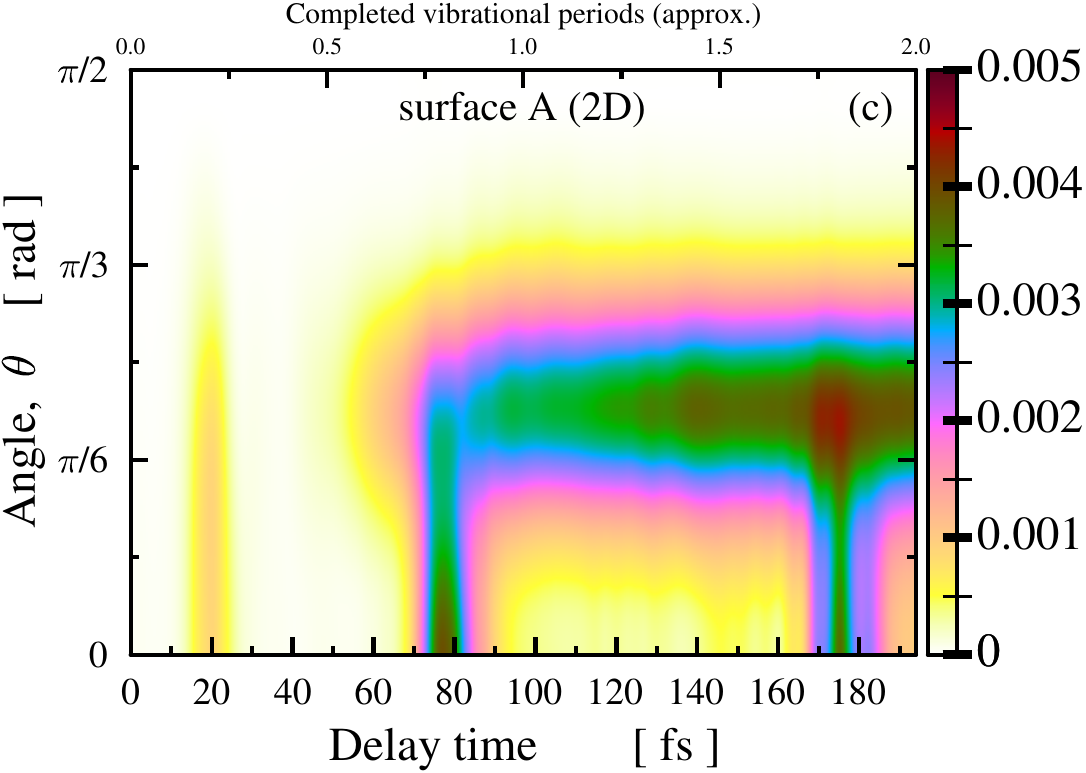}\includegraphics[width=0.48\textwidth]{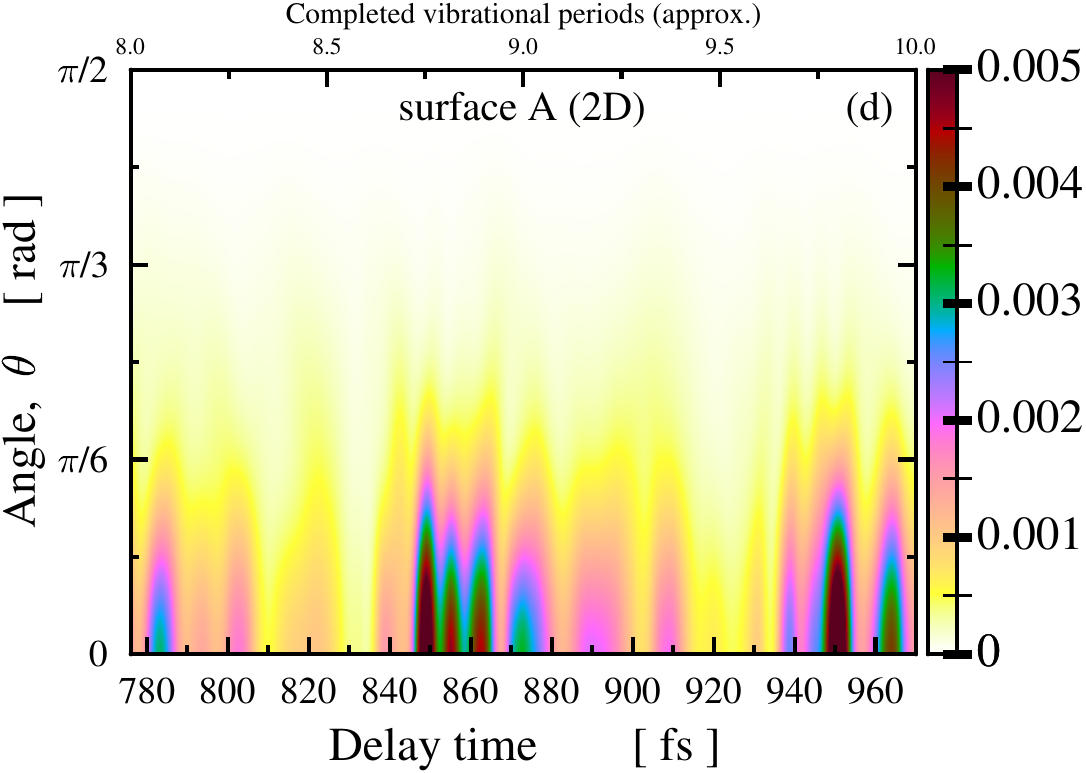}
\par\end{centering}
\caption{\label{fig:fig7}Fragment angular distribution of the dissociating
NaH molecule for the $\mathrm{V_{A}}(\mathrm{R}$) electronic state
as a function of time delay. Results are presented both for the 1D
and 2D schemes. }
\end{figure}

It can be seen in Fig. 7 that the amount of the dissociation yield
of the $\mathrm{V_{A}}(\mathrm{R})$ state is almost negligible. Both
in the 1D and 2D schemes this amount is two orders of magnitude smaller
than that of the two other states. Nevertheless, it is still worth
analyzing the structure of these panels. Again, at short delay times
the 1D and 2D results are similar. Only very small differences can
be realized in very small quantities. On the contrary, at large delay
times at around $\mathrm{t_{del}=850\,\mathrm{fs}}$ an interesting
phenomenon is experienced. The dissociation yield increases on the
$\mathrm{V_{A}}(\mathrm{R}$) state in the 2D model. The reasons might
be as follows: i)the pump pulse does not only transfer the populations
but also rotates the molecules and at a long delay time - before the
probe pulse is switched on - the molecules could rotate close to the
polarization direction of the pump pulse; ii) because two-photon process
might be initiated, the probe pulse can also transfer back some populations
from the $\mathrm{V_{X}}(\mathrm{R})$ and$\mathrm{V_{B}}(\mathrm{R})$
states to the $\mathrm{V_{A}}(\mathrm{R})$. By collecting sufficiently
large kinetic energy due to the rovibration, the dissociation yield
from the $\mathrm{V_{A}}(\mathrm{R})$ surface can increase. In this
process, due to the direction of the $\mathrm{\mu_{X,A}}$, the population
transfer from the $\mathrm{V_{X}}(\mathrm{R})$ state via the LIAC
has a more significant importance. The fingerprint of the latter effect
can clearly be seen in Fig. 9b as well, which will be further discussed.

\begin{figure}
\begin{centering}
\includegraphics[width=0.48\textwidth]{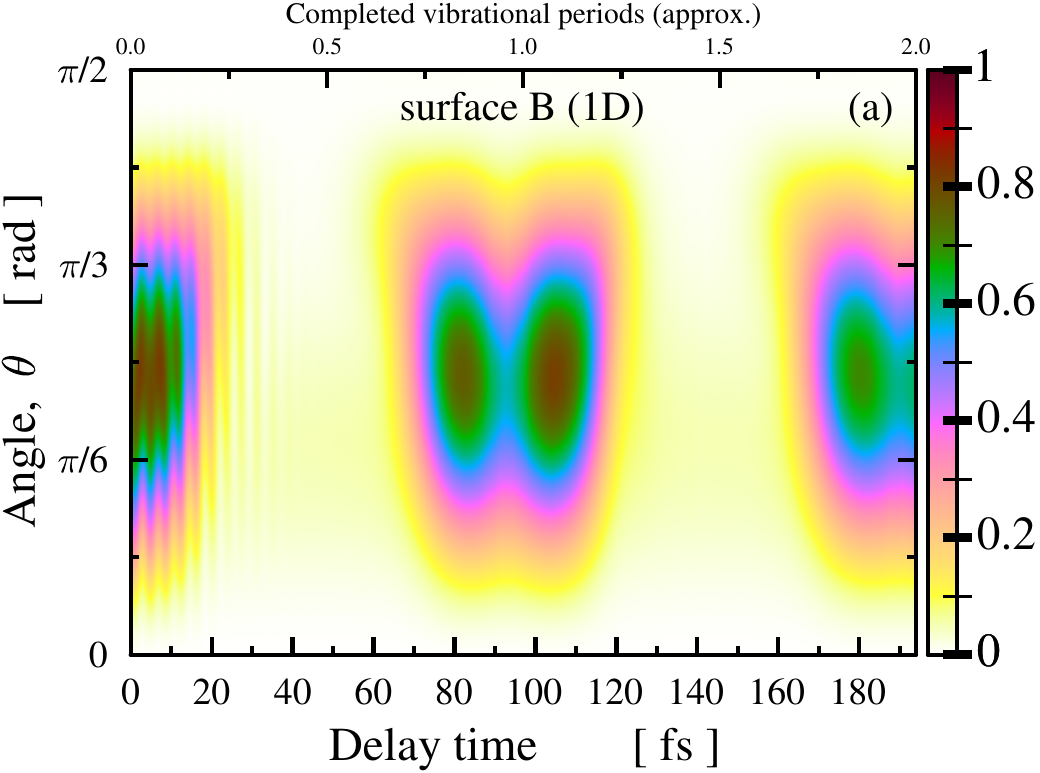}\includegraphics[width=0.48\textwidth]{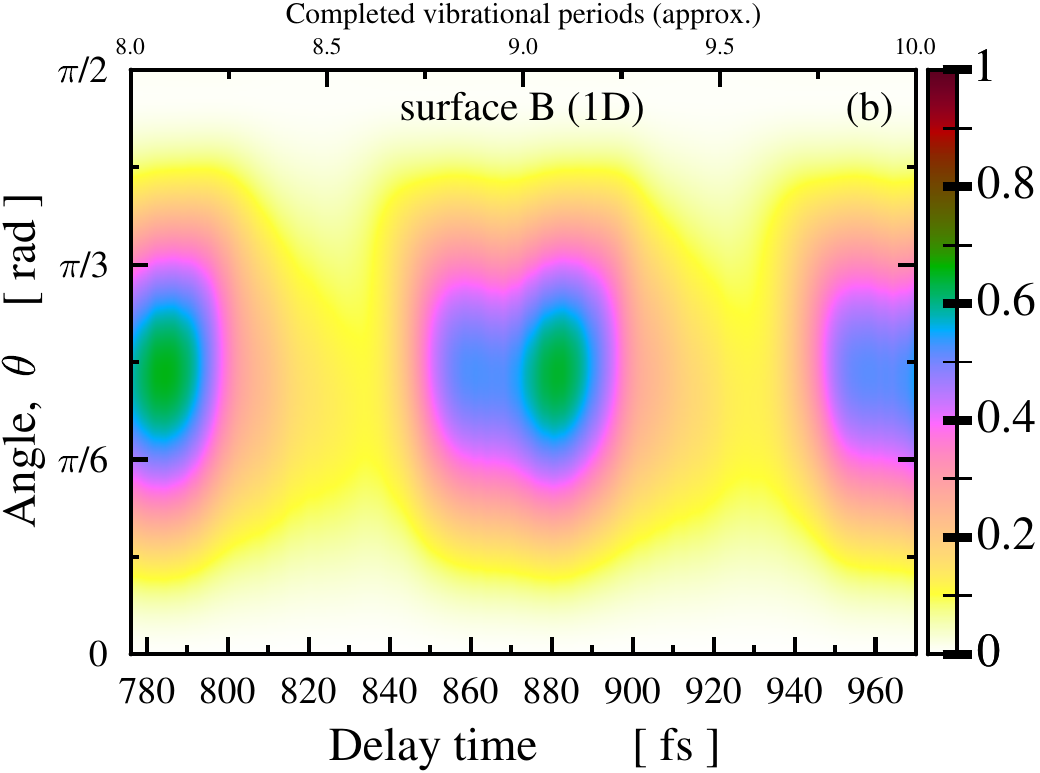}
\par\end{centering}
\begin{centering}
\includegraphics[width=0.48\textwidth]{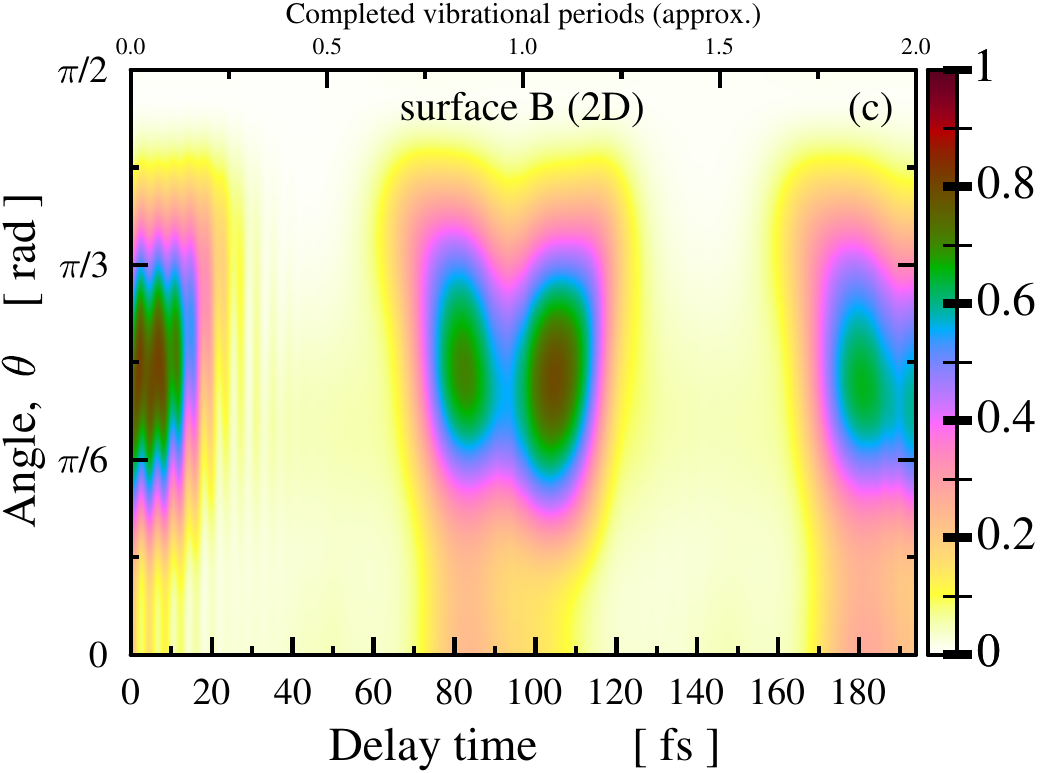}\includegraphics[width=0.48\textwidth]{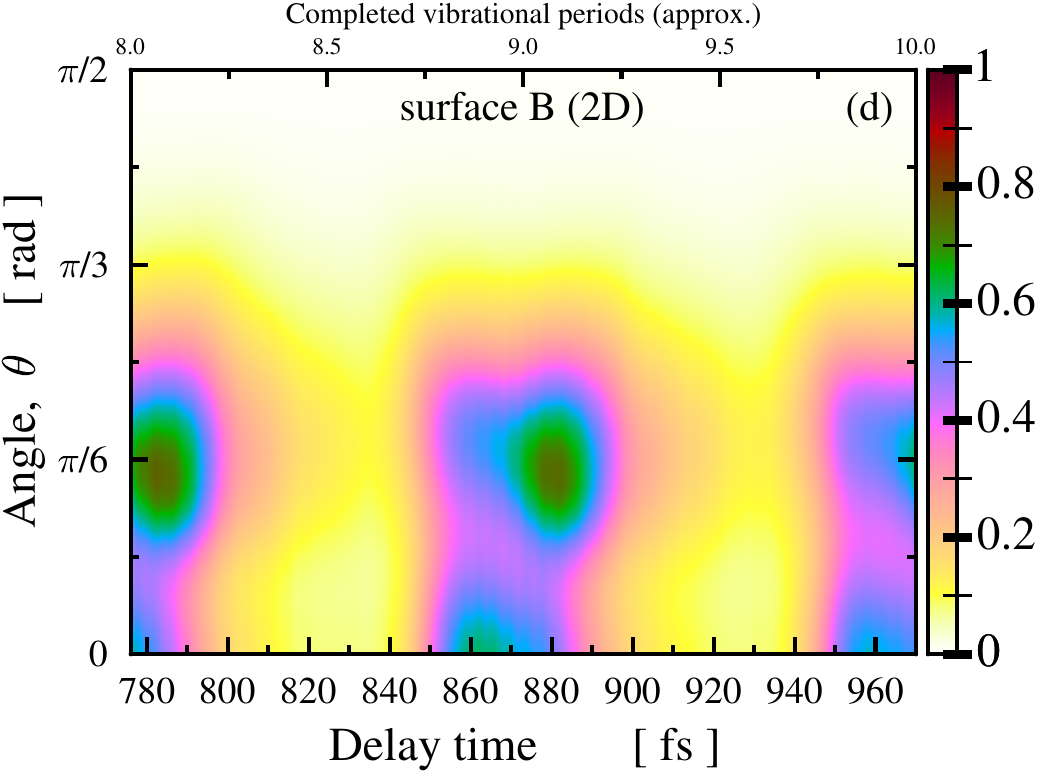}
\par\end{centering}
\caption{\label{fig:fig8}Fragment angular distribution of the dissociating
NaH molecule for the $\mathrm{V_{B}}(\mathrm{R}$) electronic state
as a function of time delay. Results are presented both for the 1D
and 2D schemes. }
\end{figure}

Fragment angular distribution of the $\mathrm{V_{B}}(\mathrm{R})$
state is displayed in Fig. 8. In the 1D scheme it is immediately apparent
that dissociation does not occur either at $\mathrm{\theta=0}$ or
at $\mathrm{\theta=90}$ degrees. At $\mathrm{\theta=90}$ degrees
the pump pulse does not transfer molecules to the $\mathrm{V_{A}}(\mathrm{R)}$
state in this direction. While at $\mathrm{\theta=0}$ the reason
for the lack of dissociation yield is that the probe pulse could not
transfer further the population from the $\mathrm{V_{A}}(\mathrm{R}$)
to $\mathrm{V_{B}}(\mathrm{R}$) as the $\mathrm{\mu_{AB}}$ is perpendicular
to the molecular axis. This tendency holds for the whole studied delay
times, because the rotation of the molecules is frozen. In short delay
times the essential difference between the 1D and 2D models is that
the dissociation appears in 2D even at $\mathrm{\theta=0}$ degree.
This effect can then become stronger, as the delay time increases.
The pump pulse does not only create an anisotropic angular distribution
of population on the $\mathrm{V_{A}}(\mathrm{R}$) state, but also
initiates a rotation which further increases this asymmetry, as the
delay time goes on. Then the probe pulse comes and with the transition
dipole moment $\mathrm{\mu_{AB}}$ which is perpendicular to the molecular
axis can further transfer the population to the $\mathrm{V_{B}}(\mathrm{R}$)
state. The transferred molecules continue to rotate further on the
$\mathrm{V_{B}}(\mathrm{R})$ state during the dissociation and finally
are aligned parallel to the polarization direction. However, this
process cannot take place in the 1D model. As the delay time increases,
and the rotation becomes more and more significant, the probe pulse
finds molecules closer and closer to the polarization direction, therefore
the larger values of the fragment angular distribution shift towards
to smaller $\mathrm{\theta}$ angles. Similarly to Figs. 6-7 discussed
before, the temporal periodicity of the fragment angular distribution
can again be interpreted as the consequence of the behavior of the
field-free nuclear wave packet (see Figure 3).

\begin{figure}
\begin{centering}
\includegraphics[width=0.48\textwidth]{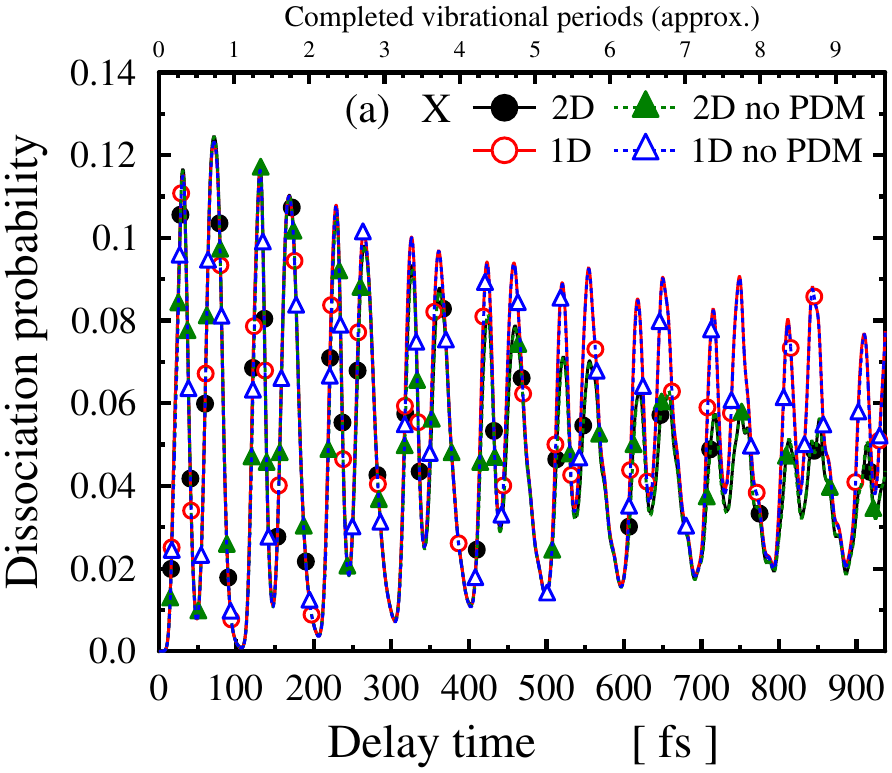}\includegraphics[width=0.48\textwidth]{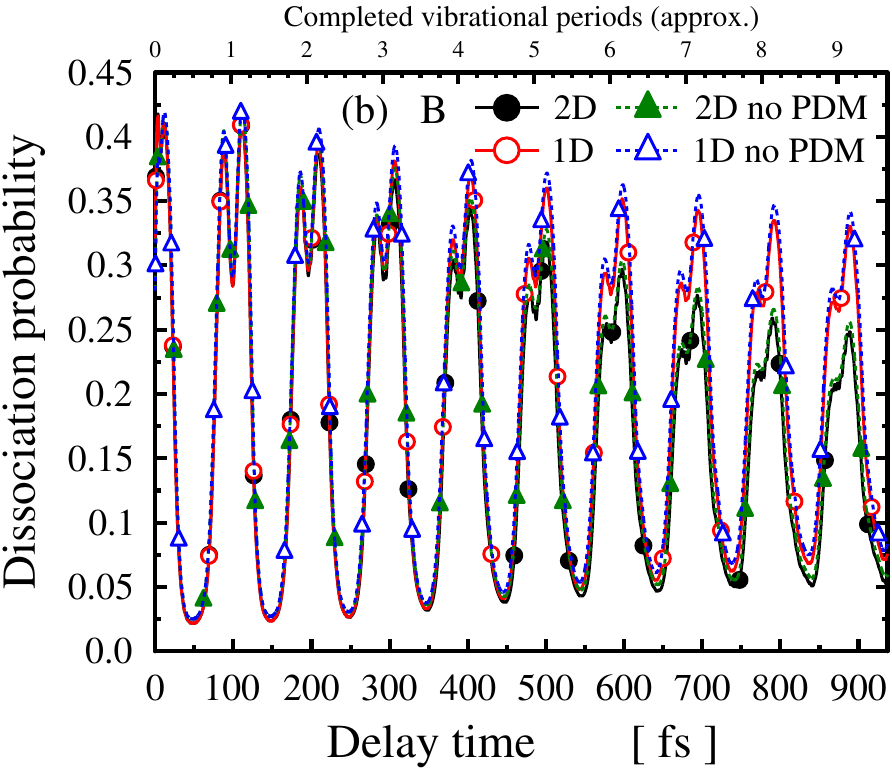}
\par\end{centering}
\begin{centering}
\includegraphics[width=0.48\textwidth]{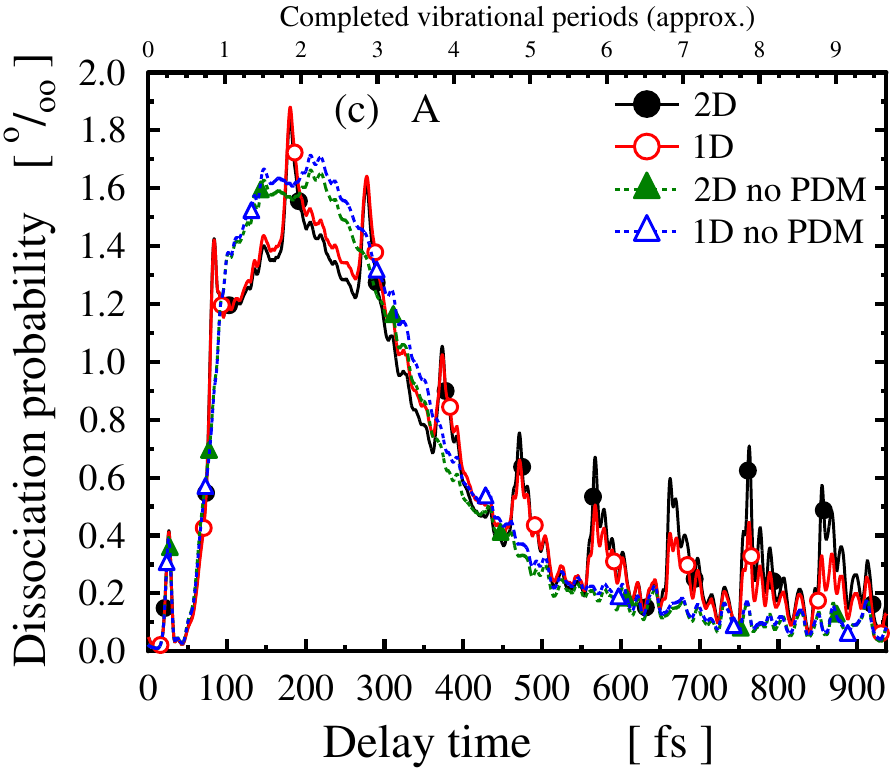}\includegraphics[width=0.48\textwidth]{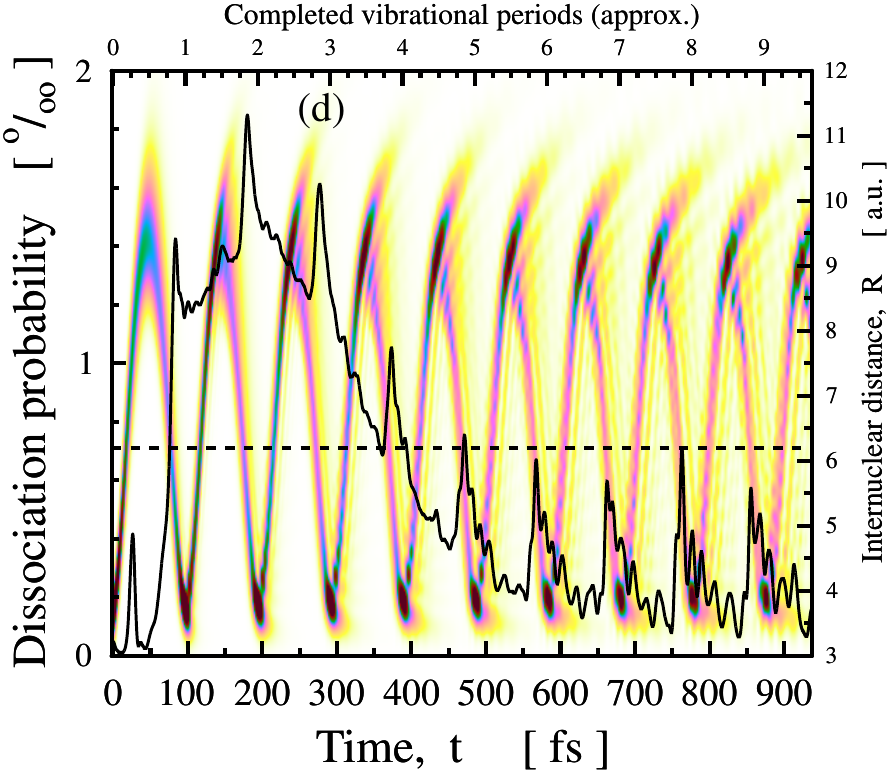}
\par\end{centering}
\caption{\label{fig:fig9} Dissociation probabilities corresponding to each
of the individual states as a function of time delay between the pump
and probe pulses. Calculations are displayed both of the 1D and 2D
schemes neglecting the permanent dipole moments (on panels (a), (b)
and (c)) as well. Shown are the results (panel (d)) for 2D dissociation
probability of the $\mathrm{V_{A}}(\mathrm{R)}$ state with PDM (scaled
in left side) and the field-free vibrational motion of the nuclear
wave packet (scaled in right-side).}
\end{figure}

Finally we discuss the role of the permanent dipole moment by means
of the investigation of the dissociation yield and the fragment angular
distribution from the $\mathrm{V_{A}}(\mathrm{R}$) state. In Fig.
9 the dissociation yields are presented with and without the permanent
dipole moment for each of the individual states both in the 1D and
2D schemes. It can be seen in Fig. 9a that independently from the
model used - on the applied scale - practically there is no difference
between the results obtained with and without using PDM for the case
of the $\mathrm{V_{X}}(\mathrm{R}$) state. This finding however,
does not completely hold for the $\mathrm{V_{B}}(\mathrm{R)}$ state
(see on panel (b)) in which minor differences can be realized between
the two different results, especially at longer delay times both in
the 1D and 2D schemes. However, inspecting Fig. 9c, an interesting
observation can be made concerning the role of the permanent dipole
moment of the $\mathrm{V_{A}}(\mathrm{R}$) state. \textcolor{black}{The
curve of the dissociation yield of state $\mathrm{V_{A}}(\mathrm{R)}$
is significantly modified by the effect of the PDM. Its shape possesses
a fairly rich structure and shows some regular spike-like peaks compared
to the simple behavior of the dissociation probability calculated
without applying the PDM. Additional numerical simulations have shown
that the background curve on the dissociation probability of surface
$\mathrm{V_{A}(R)}$ is clearly due to that some amounts of the population
are transferred back to surface $\mathrm{V_{A}(R)}$ by the probe
pulse from the dissociation region of surface $\mathrm{V_{B}(R)}$.
The rising of this curve, during the period of (50-150) fs, coincides
with that period when the wave packet on the $\mathrm{\mathrm{V_{B}(R)}}$
state arrives to the dissociative region ($\sim$ at 12 au). While
the decrease, starting at $\sim$ 250 fs, is the result of an artificial
effect, caused by the absorption of the wave packet in the CAP region
at the end of the grid. }Further analyzing the result, one can notice
that the peaks appear when the wave packet comes back from the right
side of $\mathrm{V_{A}}(\mathrm{R)}$, and gets close to the minima
of the potential energy surface. (see Fig. 9d). In order to understand
better the dissociation process happening on the $\mathrm{V_{A}}(\mathrm{R)}$
surface, several different simulations have been performed using model
PDM and TDM functions. At first, we set the value of the PDM as $\mathrm{\mu_{A}=2.5\,au}$.
The peaks do not appear in this case either, and the obtained result
is almost entirely identical to the curve calculated with $\mathrm{\mu_{A}=0\,au}$.The
position of the peaks, except for only one case (the first spike),
is determined by the delay times when the distance of the atoms approaching
each other coincides with the minimum of the $\mathrm{V_{A}}(\mathrm{R)}$
potential curve (see Fig. 9d). The size of the spikes, on the other
hand, is greatly influenced by the slope of the $\mathrm{\mu_{A}}$
function. If the permanent dipole function is steeper, the peaks of
the spikes are higher, but if the function is less steep, the peaks
of the spikes are lower. In the limiting case, if the dipole function
approaches to a constant value, the spikes disappear. 

In summary, the resulting dissociation probability of the $\mathrm{V_{A}}(\mathrm{R}$)
state is a complex effect of many factors. It is influenced by the
minimum of $\mathrm{V_{A}}(\mathrm{R}$) and its coupling to the other
two states, as well as the shape of the $\mu_{A}$ dipole moment function
etc.., but from our current study it is not clear yet what the main
underlying processes are. 

\begin{figure}
\begin{centering}
\includegraphics[width=0.48\textwidth]{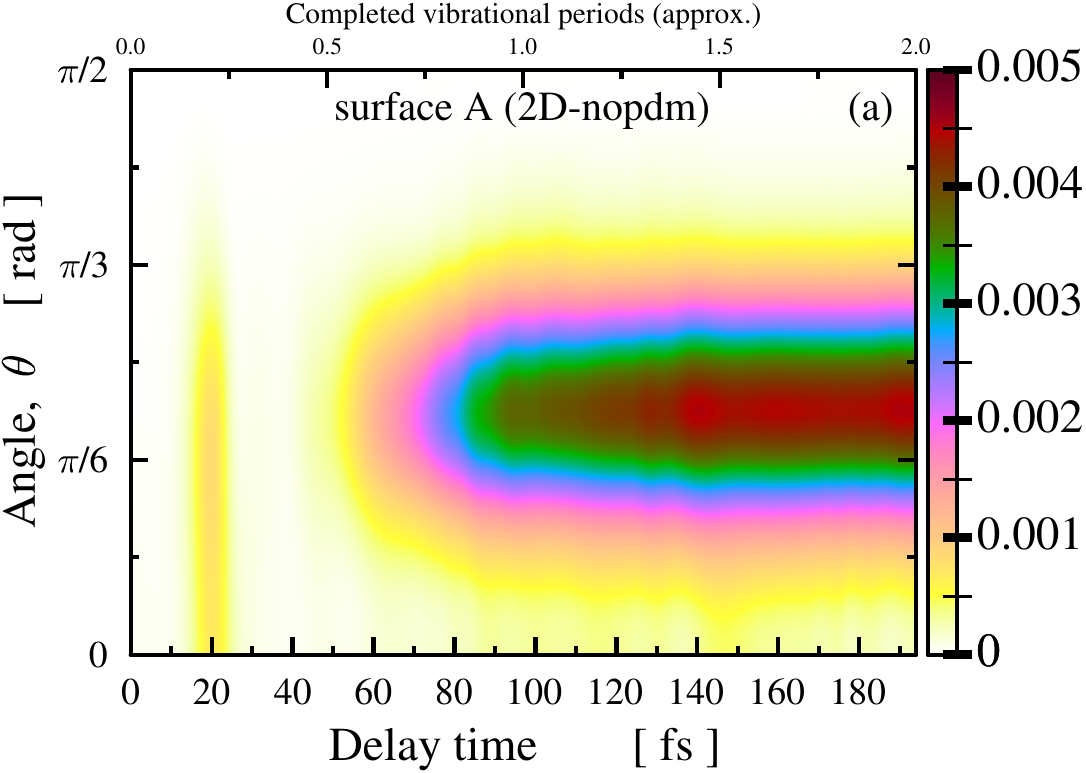}\includegraphics[width=0.48\textwidth]{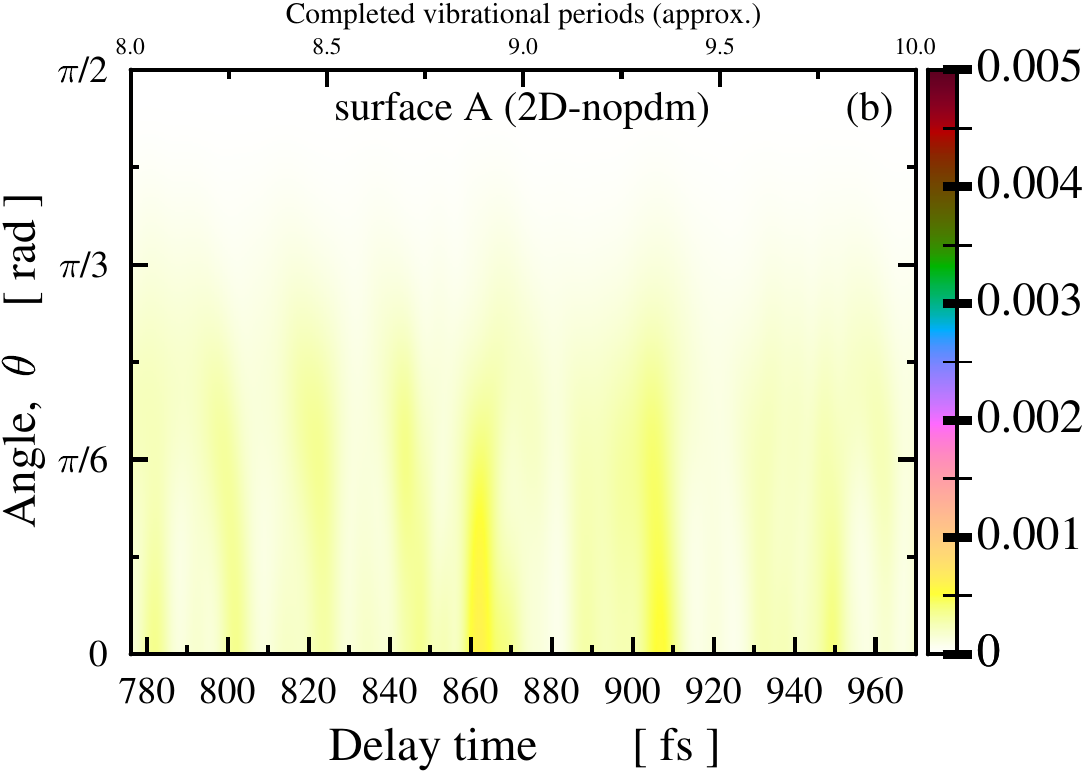}
\par\end{centering}
\begin{centering}
\includegraphics[width=0.48\textwidth]{Figures/ang-dist-2d-s2-0}\includegraphics[width=0.48\textwidth]{Figures/ang-dist-2d-s2-8}
\par\end{centering}
\caption{\label{fig:fig10} Fragment angular distribution of the dissociating
NaH molecule for the $\mathrm{V_{A}}(\mathrm{R}$) electronic state
as a function of time delay. Results are presented for the 2D model
both the cases of neglecting and including the permanent dipole moment. }
\end{figure}

In Fig. 10 the fragment angular distribution is displayed of the $\mathrm{V_{A}}(\mathrm{R)}$
state in 2D with and without using PDM. Although, as it has already
been revealed that the dissociation product of the $\mathrm{V_{A}}(\mathrm{R)}$
state is two orders of magnitude smaller than that of the $\mathrm{V_{X}}(\mathrm{R)}$
and $\mathrm{V_{B}}(\mathrm{R)}$ states, it is attention-grabbing
that the PDM significantly influences the result. It can be observed
that the dissociation yield significantly increases in the vicinity
of $\mathrm{\theta=0}$ degree which can be seen as the direct consequence
of the joint effect of the rotation as well as the LICI1 and LIAC
between the $\mathrm{V_{A}}(\mathrm{R)}$ and the two other states.
The rotation becomes even more efficient due to the effect of the
PDM and because considerable two-photon effect can be caused by the
probe pulse, larger amount of population could transfer back with
sufficient kinetic energy from the $\mathrm{V_{X}}(\mathrm{R)}$ and
$\mathrm{V_{B}}(\mathrm{R)}$ states to the $\mathrm{V_{A}}(\mathrm{R)}$
one, so as to enhance the dissociation. The spikes that appear at
around $\mathrm{\theta=0}$ degree at small delay times are also due
to the PDM and these become stronger at larger delay times due to
the increased kinetic energy gathered from the rotation.

\part*{IV.Conclusions}

In summary, we have discussed the impact of the light-induced non-adiabatic
effect as well as the molecular rotation on the dissociation dynamics
of the $\mathrm{NaH}$ molecule. The three lowest lying electronic
states were involved in the pump and probe numerical calculations.
Obtained results clearly demonstrated that the permanent dipole moments
of the molecule do not play a considerable role in the total dissociation
yield, at least for the laser parameters applied. The latter, essentially
arrives from the dissociation yields of $\mathrm{V_{X}}(\mathrm{R)}$
and $\mathrm{V_{B}}(\mathrm{R)}$ states for which the effect of PDM
is practically negligible. 

\textcolor{black}{However, the inclusion of the permanent dipole moment
of the $\mathrm{V_{A}}(\mathrm{R)}$ state resulted in a rich pattern
on the curve of the dissociation yield. The shape of the curve itself
originates from the back-transferred population to the $\mathrm{V_{A}(R)}$
state initiated by the probe pulse from the dissociation region of
surface $\mathrm{V_{B}(R)}$. The regular peaks displayed are due
to the permanent dipole moment of surface $\mathrm{V_{A}(R)}$ and
their height is strongly affected by the gradient of the $\mu_{A}$
function, apart from the initial peak. An additional spectacular trace
of these peaks appear too, in the fragment angular distribution. Nevertheless,
the underlying physical mechanism, behind of the formation of these
peaks, is not yet clear. For the in-depth further understanding and
explaining of our current findings, more detailed analysis is required.}

In the future, we plan to perform further investigations for this
system. This will include new studies using larger energies for the
probe pulse so as to create new LICIs between the $\mathrm{V_{X}}(\mathrm{R)}$
and $\mathrm{V_{B}}(\mathrm{R)}$ states, as well as applying different
pump and probe intensities to discuss properly the one and two photon
processes. To understand precisely the behavior of the dissociation
of the $\mathrm{V_{A}}(\mathrm{R)}$ state further analyses are needed
which can be the subject of the forthcoming studies. 

\section*{Acknowledgment}

The authors are indebted to NKFIH for funding (Grant No. K146096).
A.C. is grateful for the support of the János Bolyai Research Scholarship
(BO/00474/22/11) of the Hungarian Academy of Sciences. This work was
supported by the ÚNKP-23-5 New National Excellence Program of the
Ministry for Culture and Innovation from the source of the National
Research, Development and Innovation Fund.

\end{document}